\documentclass[lettersize,journal]{IEEEtran}
\usepackage{amsmath,amsfonts}
\usepackage{algorithmic}
\usepackage{algorithm}
\usepackage{array}
\usepackage[caption=false,font=normalsize,labelfont=sf,textfont=sf]{subfig}
\usepackage{textcomp}
\usepackage{stfloats}
\usepackage{url}
\usepackage{verbatim}
\usepackage{graphicx}
\usepackage{cite}
\usepackage{comment}
\usepackage{amssymb}
\usepackage{amsmath}
\usepackage{graphicx}
\usepackage{amsthm}
\usepackage{multicol}
\usepackage{balance}
\usepackage{lineno}
\theoremstyle{remark}

\newtheorem{remark}{Remark}
\newtheorem{condition}{Condition}

\newtheorem{assum}{Assumption}
\usepackage{enumitem}
\usepackage{booktabs}
\usepackage{xcolor} 
\usepackage{soul}
\sethlcolor{white}

\begin{document}

\title{Nonlinear Fault Tolerant Control of Wind Turbine Pitch Systems: A Comparative Study}

\author{Alessio Dallabona, Mogens Blanke, Dimitrios Papageorgiou
\thanks{Manuscript received October 29, 2024; revised June 10, 2025. (Corresponding author: Alessio Dallabona.)}
\thanks{All the authors are with Technical University of Denmark (DTU), 2800, Kgs. Lyngby, Denmark (email: aldall@dtu.dk; mobl@dtu.dk; dimpa@dtu.dk)}
}

\markboth{IEEE Transactions on Control Systems Technology,~Vol.~14, No.~8, May~2026}%
{Shell \MakeLowercase{\textit{et al.}}: A Sample Article Using IEEEtran.cls for IEEE Journals}


\maketitle

\begin{abstract} Electro-hydraulic pitch systems account for about one fifth of wind-turbine failures, motivating the development of resilient control strategies for fault mitigation. The study considers a large set of robust nonlinear control designs of different feedback principles and  architectures, and assesses their fault-tolerant capabilities in comparison to each other and against the industrial standard, i.e. the Proportional-Integral (PI) controller. The implementation and tuning of each scheme is based on a high-fidelity simulator to ensure nominal performance, whereas each controller is deployed unchanged on an experimental apparatus that includes a full-scale electro-hydraulic pitch system. The comparative evaluation of the considered fault-tolerant solutions is carried out through several tests that cover nominal operation and twenty-five representative fault scenarios, including sensor, process, and actuator faults. The analysis highlights the trade-off between performance preservation under faults and increased actuator chattering. The experimental results show that introducing a dynamic input-reconstruction module into the design of the nonlinear controllers significantly reduces actuator chatter without compromising their fault-tolerant properties.
\end{abstract}


\section{Introduction}\label{sec1}
\IEEEPARstart{W}{ind} turbine electricity production increased by a record 273~TWh (17\%) in 2021—the largest annual growth among all renewable power technologies \cite{ieaWindElectricityAnalysis2022}. This trend continued steadily with over 200~TWh added annually, reaching a total wind generation of 2330~TWh in 2023 \cite{IEA_Wind_2024}. Such output nearly matched the combined generation from all other non-hydro renewables. In pursuit of the Net-Zero Emissions target by 2050, the wind energy sector is moving toward larger turbines, higher power, and a growing number of offshore installations \cite{ieaWorldEnergyOutlook2022}. In this context, minimizing downtime becomes a critical objective, and reducing unplanned outages is a major concern. This is especially true for offshore wind farms, where revenue losses are amplified due to the greater complexity and longer duration of maintenance interventions compared to installations on land. 

Approximately 20\% of wind-turbine failures originate in the pitch-regulation subsystem, which, in high-power machines, is typically based on electro-hydraulic actuation \cite{gayoFinalReportSummary2011}. In practice, the pitch system is housed inside the turbine nacelle and hub assembly, where one actuator per blade converts linear hydraulic motion into blade rotation about its longitudinal axis to regulate aerodynamic torque and rotor speed \cite{lanReviewFaultDiagnosis2024a}. Faults and other degradation in this mechanism directly translate into pitch imbalance and uneven aerodynamic loading. The consequent high structural loads make the operation unsafe \cite{etemaddarResponseAnalysisComparison2016a}. Hence, implementing fault detection and accommodation schemes within this subsystem can significantly enhance turbine availability. Notably, the same class of electro-hydraulic actuators (EHAs) is employed across various domains---including robotics, mobile platforms, and industrial presses---underscoring the broader relevance of the proposed control strategies.

\begin{figure*}[t]
    \centering
    \includegraphics[width=0.9\textwidth]{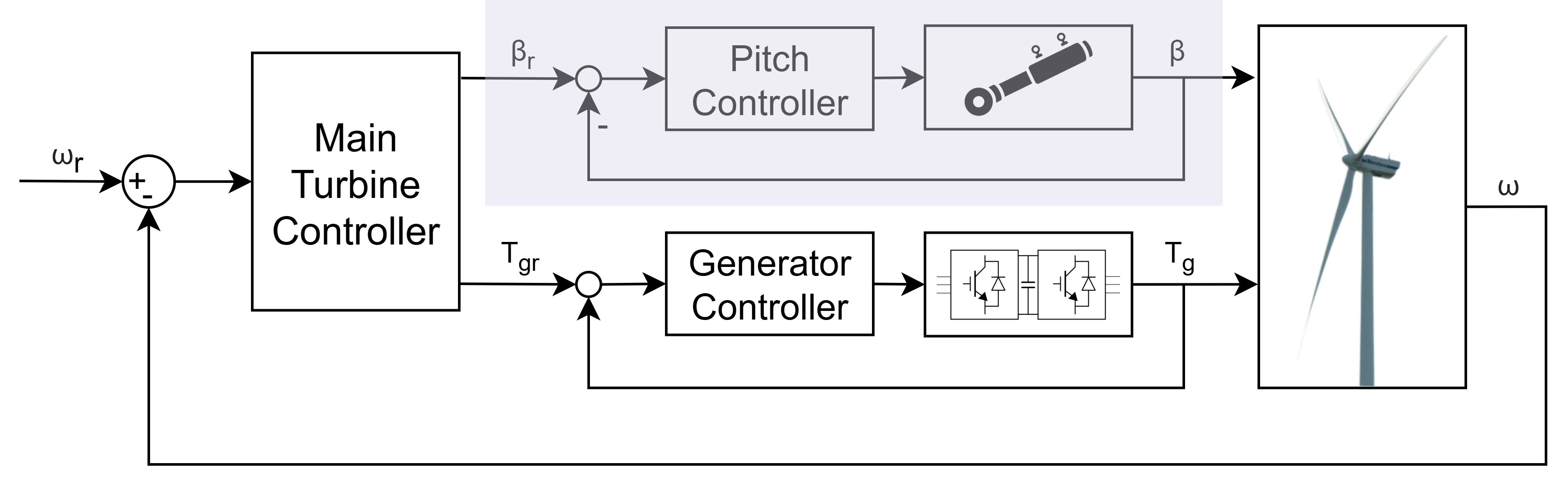}
    \caption{Block diagram of the full wind turbine control system. Main controller generates setpoints for pitch angle and generator torque. Subsystems comprise internal control loop(s) to track references. The highlighted pitch subsystem is the focus of this article.}
    \label{fig:wt_control_block}
\end{figure*}

Advanced control methodologies are key to enhancing the performance and robustness of the pitch control loop, particularly under conditions with external disturbances and faults. For reasons of compatibility with present control architectures, the focus is put on fault mitigation that is achieved without requiring explicit identification or diagnosis of the fault itself, a strategy that is referred to as \emph{passive fault-tolerant control} (PFTC). For EHA-based systems, the inherent nonlinear dynamics demand the application of nonlinear control theory to ensure global stability, rather than relying solely on local linear approximations.

Other literature on fault-tolerant control for wind turbines has primarily focused on high-level supervisory control, to govern generator torque and pitch angle references and maximize power extraction and limit mechanical loads in case of defects. An overview is provided in \cite{fekihFaultDiagnosisFault2022a,aminApplicationsFaulttolerantControl2025}, and more results can be found in \cite{chouiekhActiveFaultTolerantMPPT2025,lamModelFreeAdaptiveFaultTolerant2024,liuRobustBladePitch2024,serranoIntegralSlidingModeFaulttolerant2023e}. Only a few studies have addressed the implementation of such techniques within the \emph{internal pitch control loop} that is responsible for actuating the blade pitch angle based on the reference signal from the main controller \cite{lanFaulttolerantWindTurbine2018b,huAdaptiveFaulttolerantPitch2023a,venkaiahRotaryActuatedWindTurbine2025}. None of the aforementioned studies deals with pitch systems driven by linear electro-hydraulic actuators. While fault-tolerant strategies have been sparsely explored for generic hydraulic actuators \cite{karpenkoQuantitativeFaultTolerant2010a,wangActiveFaulttolerantControl2024,stojanovicFaulttolerantControlHydraulic2023}, a rich body of literature exists on nonlinear control techniques for EHAs, which are not explicitly designed for fault scenarios, but offer inherent robustness that can help mitigate the impact of faults. Examples can be found both for wind turbine pitch control \cite{venkaiahHarrisHawksOptimization2024, wangLoadPredictionControl2025,dallabonaDynamicInputMapping2025} and for generic EHAs \cite{manganasBacksteppingBasedController2023,komstaIntegralSlidingMode2013,kochObserverbasedSlidingMode2016,estradaSupertwistingBasedSliding2024,yamamotoPositionControllerHydraulic2025,dallabonaModelfreeSlidingMode2024,chouxAdaptiveBacksteppingControl2010a}.

\vspace{0.5em}

The objective of this article is to provide a systematic, comparative analysis of nonlinear and robust control strategies for passive fault tolerance in wind-turbine pitch systems. The goal is to identify the most suitable candidate to serve as a passive fault-tolerant baseline controller\hl{, rejecting the effects of faults in the same way that a robust controller rejects disturbances,} and to identify the limitations of the considered baseline solutions.

The controllers considered in this work are formulated within a common nonlinear pitch-system framework by building on existing nonlinear and robust-control methods to define a set of candidate fault-tolerant control structures for the hydraulic pitch system. Their formulation, implementation requirements, tuning, and fault-oriented performance are then assessed through a unified evaluation procedure, including experimental validation on a full-scale setup. This enables a consistent comparison across the resulting baseline solutions. The contributions of this work are:
\begin{enumerate}
    \item \hl{Implementation and experimental verification on an industrial-scale hydraulic pitch system of state-of-the-art and newly introduced passive fault-tolerant control strategies. These include combinations of Super Twisting Sliding Mode schemes with dynamic input mapping inversion in a position-force cascaded architecture;}
    \item \hl{Stability analysis of the closed-loop system for the different control schemes including the effects of faults;}
    \item \hl{Definition of a quantitative fault-tolerant performance evaluation framework;}
    \item \hl{Systematic comparative analysis of the proposed controllers and selection of the most suitable candidate for fault-tolerant operation including consideration on the methods' limitations.}
\end{enumerate}

The remainder of the paper is organized as follows. Section~\ref{sec2} introduces the wind turbine pitch system case study and formulates the passive fault-tolerant control problem. Section~\ref{sec3} describes the design of the controller set, and Section~\ref{sec4} analyzes their stability properties. Section~\ref{sec5} presents the high-fidelity simulator and experimental setup, while Section~\ref{sec6} reports the simulation and experimental results. Finally, conclusions and directions for future research are offered in Section~\ref{sec7}. 
 
\section{System and problem formulation}\label{sec2}

Wind turbine control aims to regulate rotor speed and, consequently, the power extracted from the wind and converted into electrical energy. A high-level block diagram of a wind turbine control architecture is illustrated in Figure~\ref{fig:wt_control_block}, with emphasis on the pitch system. Traditionally, most control design efforts focus on the high-level controller responsible for computing the reference pitch angle $\beta_{r}$ \cite{novaesmenezesReviewWindTurbine2018}. The internal pitch system is then typically approximated as a second-order stable subsystem. In this work, it is modeled in full detail, capturing all its internal dynamics and nonlinearities.

\begin{figure}[tp]
    \centering
    \includegraphics[width=\columnwidth]{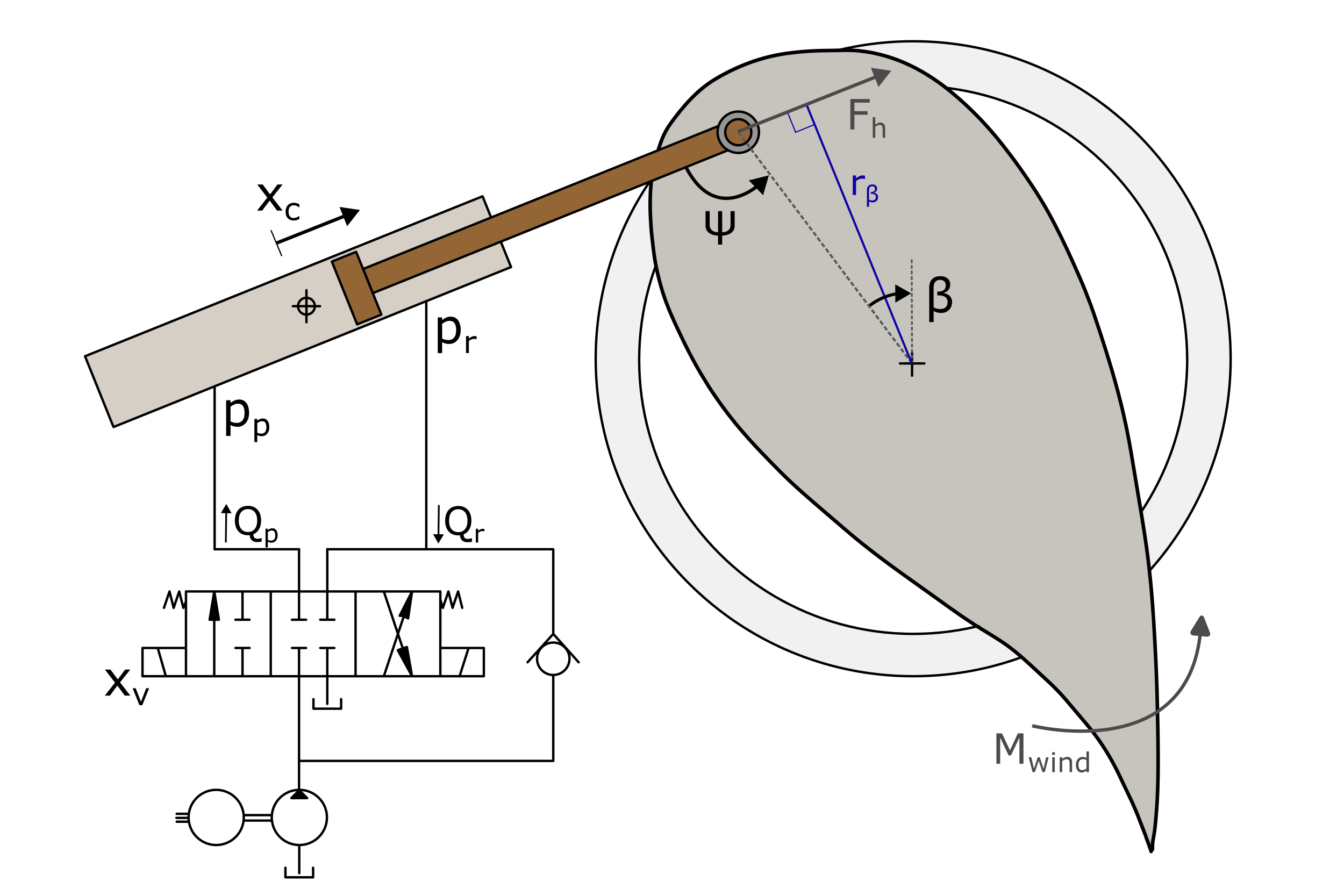}
    \caption{Schematic of a generic electro-hydraulic actuator for pitch control: cylinder and piston with blade bearing (top), DCV and oil supply circuit (bottom).}
    \label{fig:hyd_sys}
\end{figure}

The most common configuration for pitch actuation in high-power wind turbines is shown in Figure~\ref{fig:hyd_sys}. The pitch angle $\beta$ is varied by extending or retracting a hydraulic cylinder, actuated via a directional control valve (DCV). The pressures $p_p$ and $p_r$ in the piston-side and rod-side chambers of the cylinder are regulated by controlling the valve spool position $x_v$ (and velocity $v_v$), thereby determining the piston position $x_c$ and velocity $v_c$. The resulting nonlinear model of the pitch system is adopted from~\cite{komstaIntegralSlidingMode2013, dallabonaFaultDiagnosisPrognosis2025, bakhshandeRobustControlApproach2017}:
\begin{equation}\label{eq:gen_dyn}
\begin{aligned}
& J \dot{v}_{\beta}=(\overbrace{A_p p_p-A_r p_r}^{F_h}-F_{Fr,c})\,r_{\beta}(x_c) -M_{Fr,b}-M_{ext} \\
& \dot{\beta}=v_{\beta} \\
& \dot{p}_p=\frac{B}{V_p(x_c)}\left[Q_p(x_v)-A_p v_c \textcolor{blue}{- Q_{le,p} - Q_{li} + Q_{lp}}\right] \\
& \dot{p}_r=\frac{B}{V_r(x_c)}\left[-Q_r(x_v)+A_r v_c \textcolor{blue}{- Q_{le,r} + Q_{li}- Q_{lr}}\right] \\
& \dot{v}_v+2 \xi \omega_0 v_v+\omega_0^2 x_v=\omega_0^2 u_x\\
& \dot{x}_v = v_v,
\end{aligned}
\end{equation}

Here, $J$ is the rotational inertia, $A_p$ and $A_r$ are the piston and rod-side areas, $M_{Fr,b}$ and $M_{ext}$ represent friction in the bearing and external wind load, $F_{Fr,c}$ is the friction inside the cylinder and $r_\beta$ the moment arm, and $B$ is the effective bulk modulus of the fluid in each chamber. $F_h \triangleq A_p p_p-A_r p_r$ is the hydraulic force in the cylinder, and the cylinder position $x_c$ is obtained via the following static transformation on $\beta$:
\begin{equation}
    x_c = \sqrt{-\cos\big(\beta + \Phi_\mathrm{off}\big) \, 2 l_2 l_3 + l_2^2 + l_3^2} - l_1
\end{equation}
where $l_1$, $l_2$, and $l_3$ are lengths of the mechanical links in the pitch bearing assembly, and $\Phi_\mathrm{off}$ is the geometric offset angle between the actuator axis and the angle of the blade \cite{linigerDesignReliableFluid2018}.

The friction terms are modeled as:
\begin{equation} \label{eq:friction}
    \begin{aligned}
        M_{Fr,b} &= F_{c,b}\,\text{tanh}(\gamma_t\, v_\beta) + B_{v,b} v_\beta\\
        F_{Fr,c} &= F_{c,c}\,\text{tanh}(\gamma_t\,v_c) + B_{v,c} v_c,
    \end{aligned}
\end{equation}
where $F_{c,i}$ and $B_{v,i}$ are, respectively, static and viscous friction coefficients, and $\gamma_t$ a coefficient determining the shape of the $\text{tanh}$ approximation to the sign function.

The chamber volumes are defined as:
\begin{equation}\label{eq:volumes}
\begin{aligned}
& V_p(x_c) = V_{0, p}+A_p x_c\\
& V_r(x_c) = V_{0, r}+A_r\left(x_{c, m}-x_c\right).
\end{aligned}
\end{equation}

Flows are given by
\begin{equation}\label{eq:flows}
\begin{aligned}
& Q_p(x_v) = K_v(x_v)\,\delta_p\\
& Q_r(x_v) = K_v(x_v)\,\delta_r;
\end{aligned}
\end{equation}
where
\begin{equation}\label{eq:flows_pres}
\begin{aligned}
& \delta_p =  \sqrt{p_s-p_p}  \,H\left(x_v\right) +  \sqrt{p_p-p_t}\,H\left(-x_v\right)\\
& \delta_r =  \sqrt{p_r-p_t}  \,H\left(x_v\right) +  \sqrt{p_s-p_r}\,H\left(-x_v\right),
\end{aligned}
\end{equation}
the valve flow coefficient $K_v(x_v)$ is a nonlinear function of the spool position, and $p_s$ and $p_t$ are the supply and tank pressures respectively. $H(\cdot)$ is a smooth approximation to the Heaviside step function,
\begin{equation}\label{eq:smooth_H}
        H(\cdot) \approx \frac{1}{2}\left(1 + \tanh(\gamma_L\; \cdot)\right)
\end{equation}
with $\gamma_L>0$ sufficiently large. This smooth approximation will be required later to ensure Lipschitz continuity of $H$, which is necessary in the subsequent stability analysis.

The flows highlighted in blue in Eq.~\ref{eq:gen_dyn} represent leakage flows, which occur only under fault conditions. They are modeled as:
\begin{equation}
\begin{alignedat}{2}  
  Q_{le,p} &= c_{le,p}(p_p - p_t) &\qquad Q_{lp} &= c_{lp}(p_s - p_p)\\
  Q_{le,r} &= c_{le,r}(p_r - p_t) &\qquad Q_{lr} &= c_{lr}(p_r - p_t)\\
  Q_{li}   &= c_{li}(p_p - p_r)
\end{alignedat}
\end{equation}
with all proportional coefficients equal to zero during nominal operation.

\begin{table}[h!]
    \centering
    \caption{Summary of considered fault scenarios affecting the pitch system. 
    Unless otherwise specified, all quantities are expressed in SI units. Percentages denote remaining value relative to nominal (100\%).}

    \label{tab:fault_scenarios}
    \begin{tabular}{@{}lcc@{}}
        \toprule
        \textbf{Fault type} & \textbf{Value} & \textbf{Cat.} \\
        \midrule
        Bias ($\beta$ sensor)               & $0.01$               & 1 \\
        Gain ($\beta$ sensor)               & $0.93$               & 1 \\
        Bias (p$_\mathrm{p}$ sensor)  & $1.5\times10^{6}$    & 2 \\
        Gain (p$_\mathrm{p}$ sensor)  & $0.85$               & 2 \\
        Bias (p$_\mathrm{r}$ sensor)  & $5\times10^{5}$      & 2 \\
        Gain (p$_\mathrm{r}$ sensor)  & $0.6$                & 2 \\
        Bias (p$_\mathrm{s}$ sensor)  & $-5\times10^{5}$     & 2 \\
        Gain (p$_\mathrm{s}$ sensor)  & $0.95$               & 2 \\
        Bulk modulus B reduction        & $60\%$                & 3 \\
        External leak in p $c_{le,p}$     & $0.5\times10^{-11}$  & 3 \\
        External leak in r $c_{le,r}$    & $3\times10^{-11}$    & 3 \\
        Internal leak $c_{li}$          & $8\times10^{-11}$    & 3 \\
        Friction gain $\Delta B_{v,c}$     & $100$                & 3 \\
        Friction gain $\Delta F_{c,c}$     & $50$                 & 3 \\
        Friction gain $\Delta B_{v,b}$  & $100$                & 3 \\
        Friction gain $\Delta F_{c,b}$  & $20$                 & 3 \\
        Valve coefficient bias $\Delta k_{v,p}$       & $-0.1\times10^{-7}$   & 3 \\
        Valve coefficient bias $\Delta k_{v,r}$        & $-0.4\times10^{-6}$   & 3 \\
        Valve leak in p $c_{lp}$     & $1\times10^{-7}$     & 3 \\
        Valve leak in r $c_{lr}$     & $0.5\times10^{-6}$   & 3 \\
        System pressure drop $\Delta p_s$     & $-0.2\times10^{6}$   & 3 \\
        DCV decreased damping $\zeta$     & $50\%$                  & 4 \\
        DCV decreased natural freq.\ $\omega_0$ & $ 83 \%$           & 4 \\
        Actuator bias $u$              & $0.3$                & 4 \\
        Actuator gain $u$              & $1.15$               & 4 \\
        \bottomrule
    \end{tabular}%
\end{table}

The input dynamics of the DCV is commonly modeled in the literature as a second-order system, with natural frequency $\omega_0$ and damping ratio $\xi$, to capture the internal loop behavior typical of modern valves. The normalized control input $u_x\in[-1,1]$ represents the reference spool position. As one can verify in the presented literature, a commonly adopted assumption is to neglect the valve dynamics entirely and thus treat $x_v$ as the direct control input. Further details are provided in \cite{dallabonaDynamicInputMapping2025}. Under this assumption, $x_v$ is instantaneously set to any desired value, i.e.:
\begin{equation}
    u \triangleq x_v, \quad \text{and} \quad Q_p(x_v), Q_r(x_v) \equiv Q_p(u), Q_r(u),
\end{equation}
which will be used throughout the remainder of this work. The impact of this approximation during control design depends on the valve’s dynamic performance, as neglecting its dynamics assumes the reference position is reached instantaneously, whereas a small delay is in fact present. The feasibility of this simplification will be assessed through simulations and experimental tests to verify whether neglecting the valve dynamics during control design still yields satisfactory performance when applied to the actual system. Moreover, the valve flow coefficient---usually available in the form of a lookup table---is approximated by a $9^{\text{th}}$-order polynomial function of the input,
i.e. $K_v(u) \triangleq \sum\limits_{i=0}^9 \alpha_i u^i$, $\alpha_i \in \mathbb{R}$,
to obtain a smooth and analytically differentiable mapping required during the control design stage.

The control objective is to ensure that the pitch angle $\beta$ tracks a smooth reference signal $\beta_r\in \mathcal{C}^3$---generated by an outer loop, see Section 5---by minimizing the tracking error
\begin{equation}\label{eq:error_def}
e_\beta = \beta - \beta_r
\end{equation}
such that $e_\beta \rightarrow 0$. While the controllers are not explicitly designed to account for fault tolerance, they are evaluated in the context of PFTC to assess the level of inherent robustness and fault resilience they provide under fault conditions.

A comprehensive summary of potential faults affecting electro-hydraulic pitch systems is presented in~\cite{dallabonaFaultDiagnosisPrognosis2025}, including their severity and impact on safe operation. To enable a comprehensive assessment of fault-tolerant capabilities, most of these fault scenarios are incorporated into the analysis, including friction increase in the bearing as a consequence of accounting for the pitching dynamics. The fault types and their values are summarized in Table~\ref{tab:fault_scenarios} and are selected to be representative, inducing a non-negligible effect on the closed-loop response and enabling a meaningful assessment of fault-rejection capabilities\hl{, considering only one fault at a time.}

\hl{The individually tested faults are then grouped into categories to facilitate comparison of fault-rejection performance across fault classes.} The conventional grouping into sensor, process, and actuator faults is adopted. The numerical category labels and their descriptions are given in Table~\ref{tab:fault_categories}.

\begin{table}[bp]
    \centering
    \caption{Mapping of fault category numbers to descriptive names.}
    \label{tab:fault_categories}
    \begin{tabular}{@{}lc@{}}
        \toprule
        \textbf{Fault Category} & \textbf{Category Description} \\
        \midrule
        1 & Angle Sensor Faults \\
        2 & Other Sensor Faults \\
        3 & Process Faults \\
        4 & Actuator Faults \\
        \bottomrule
    \end{tabular}
\end{table}

\section{Control design}\label{sec3}

\begin{figure}[tp]
    \centering
\includegraphics[width=0.95\columnwidth]{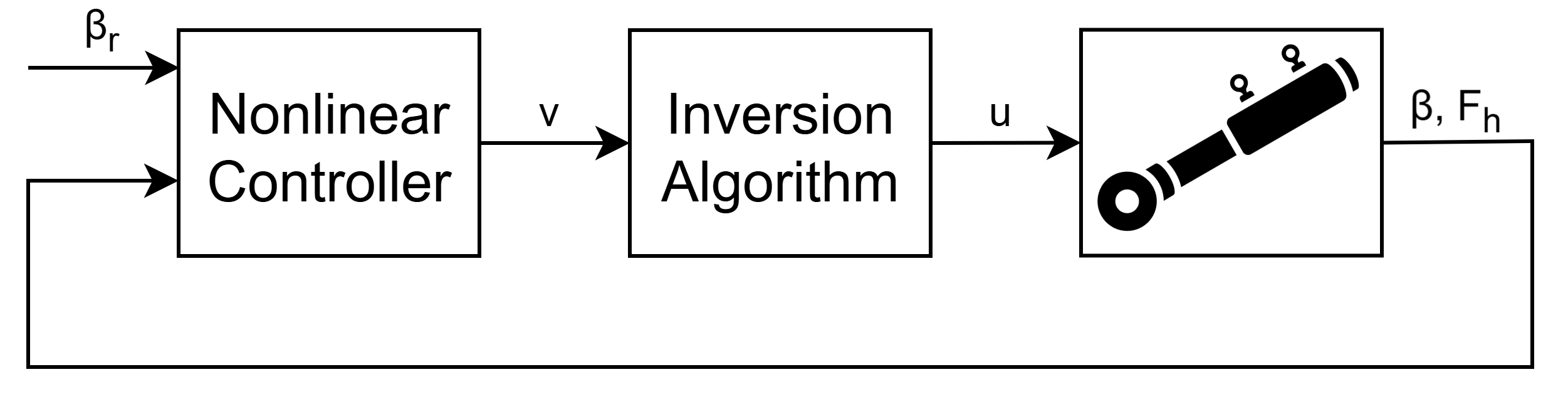}
    \caption{\hl{General cascaded architecture} used for all proposed controllers: nonlinear control block delivering a virtual input, followed by an inversion block which determines the associated control action. It corresponds to the highlighted area in Fig. \ref{fig:wt_control_block}.}
    \label{fig:control_main}
\end{figure}

All control strategies presented in this work adopt a common cascaded architecture, depicted in Figure~\ref{fig:control_main}. The structure builds upon the definition of a virtual control input $v$, which depends on the actual control input $u$ via a nonlinear relationship. This virtual input should be thought of as the controlled flow (scaled by the volume) that generates appropriate pressure to effectively actuate the pitch system. In comparison, the actual control input $u$ is the valve position that allows the generation of the necessary flow. Due to the strong nonlinearities in the expression for $v$, a closed-form inverse formula that calculates $u$ for a given $v$ is very difficult if at all possible to obtain. For this reason, the \emph{Inversion Block} is responsible for computing the appropriate value of $u$ that reflects the desired $v$. The details of the inversion block and the associated algorithm will be presented in Sec. \ref{sec:inversion} further below.

\begin{figure*}[tp]
    \centering
        \subfloat[Full-model design architecture]{%
        \includegraphics[width=0.355\textwidth]{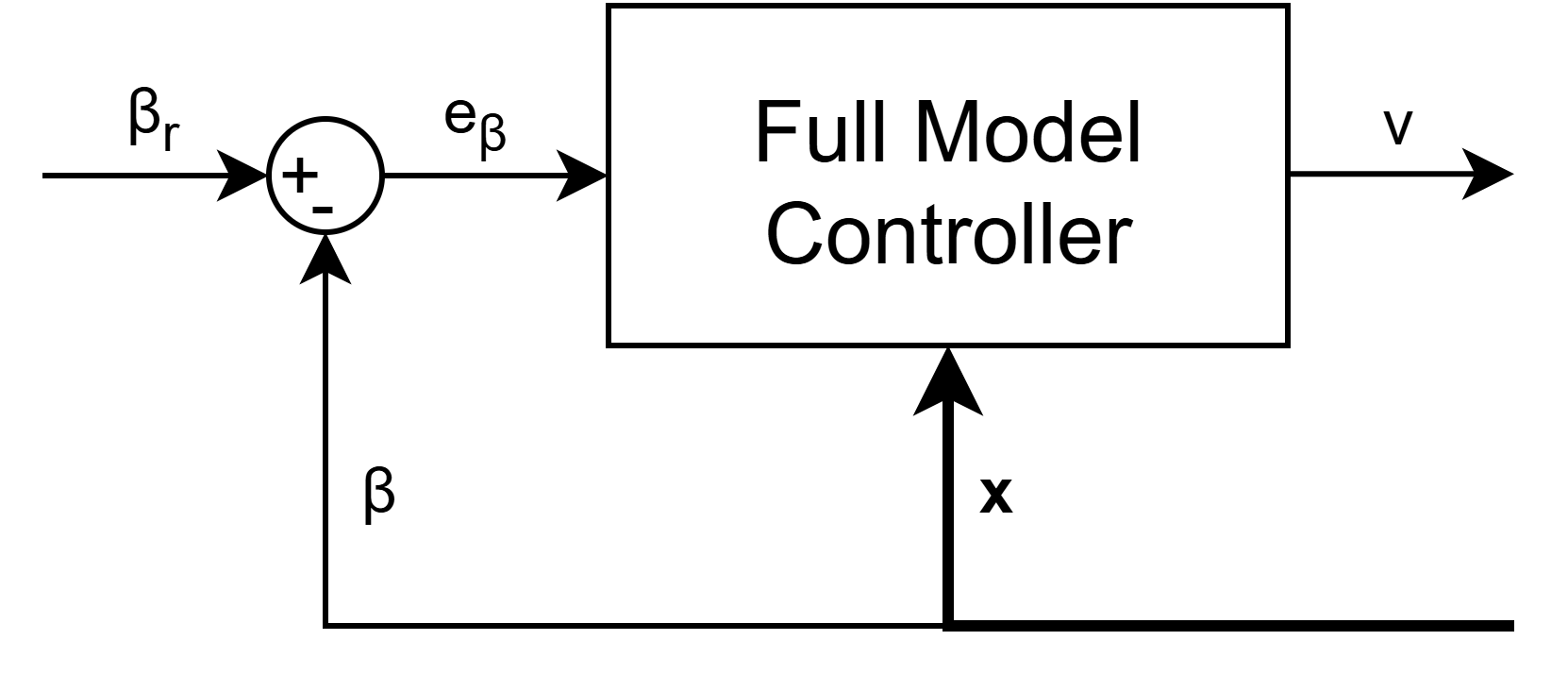}
        \label{fig:control_design_fm}
        }
    \hfill 
    \subfloat[Position–force cascade (PFC) architecture]{%
        \includegraphics[width=0.625\textwidth]{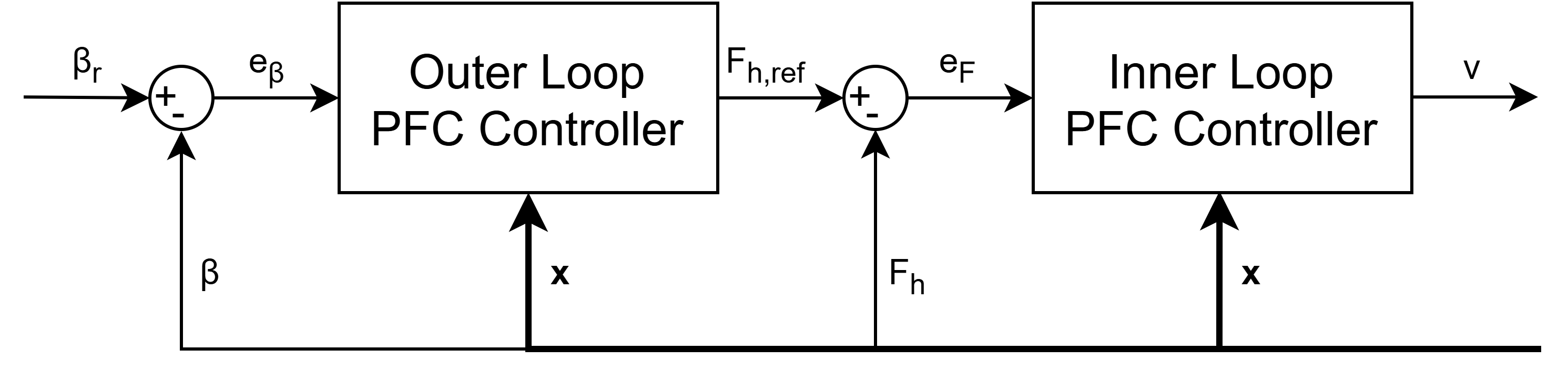}
        \label{fig:control_design_pfc}
    }
    \caption{Two configurations used for the design of the \emph{Nonlinear Control} block, introduced in Fig. \ref{fig:control_main}.}
    \label{fig:control_designs}
\end{figure*}

The \emph{Nonlinear Control} block itself can be structured in two distinct ways, as illustrated in Figure~\ref{fig:control_designs}:
\begin{enumerate}
    \item A \textit{full-model (FM) design} (Fig.~\ref{fig:control_designs}a), where a single controller is designed to act on the complete nonlinear system.
    \item A \textit{position–force cascade (PFC)} structure (Fig.~\ref{fig:control_designs}b), motivated by the mechanical-hydraulic coupling of the system. Here, the outer loop computes the force needed to track the pitch angle reference, and the inner loop regulates the hydraulic force to match this reference. This approach simplifies the design by reducing the relative degree of each subsystem.
\end{enumerate}

Each architecture is tested with a set of control strategies, summarized below:
\begin{itemize}
    \item \textbf{Full-model design:} PI, Feedback Linearization (FL), Super-Twisting Sliding Mode Control (STSMC);
    \item \textbf{PFC – Outer loop:} P, FL;
    \item \textbf{PFC – Inner loop:} PI, FL, STSMC.
\end{itemize}

The reduced number of control strategies in the outer loop of the PFC is due to the need for smooth control actions. Sharp variations in the output signal of the outer loop, $F_{h,\text{ref}}$, can cause instability or poor performance of the hydraulic inner loop, due to peaking or saturation effects. The complete list of  controllers is summarized in Table~\ref{tab:controllers}.

\subsection{Nonlinear Controller: Full-Model Design}

The full-model design involves a relative degree of three from the input $u$ to the tracking error $e_\beta = \beta - \beta_r$.

\begin{table}[tp]
    \small
    \centering
    \caption{Acronyms for controllers }
    \begin{tabular}{@{}lp{0.6\linewidth}@{}}
        \toprule
        \textbf{Controller} & \textbf{Description} \\
        \midrule
        PI & Full Model, PI controller\\
        FL & Full Model, FL controller\\
        ST & Full Model, STSMC\\
        MIFL & Full Model, FL controller with dynamic input inversion\\
        MIST & Full Model, STSMC with dynamic input inversion\\ \hline
        PFC P-PI & PFC with P controller in the outer loop and PI controller in the inner loop \\
        PFC P-FL & PFC with P controller in the outer loop and FL controller in the inner loop\\
        PFC P-ST & PFC with P controller in the outer loop and STSMC in the inner loop\\
        PFC FL-FL & PFC with FL controller in the outer loop and FL controller in the inner loop\\
        PFC FL-ST & PFC with FL controller in the outer loop and STSMC in the inner loop \\
        PFC P-MIFL & PFC with P controller in the outer loop and FL controller with dynamic input inversion in the inner loop \\
        PFC P-MIST & PFC with P controller in the outer loop and STSMC with dynamic input inversion in the inner loop  \\
        PFC FL-MIFL &  PFC with FL controller in the outer loop and FL controller with dynamic input inversion in the inner loop \\
        PFC FL-MIST &  PFC with FL controller in the outer loop and STSMC with dynamic input inversion in the inner loop \\
        \bottomrule
    \end{tabular}
    \label{tab:controllers}
\end{table}

The third time derivative of the tracking error can be written as:

\begin{equation}\label{eq:ddde}
\begin{aligned}
e_\beta^{(3)} =  &-\frac{\dot{M}_{Fr,b} + \dot{M}_{\text{ext}}}{J} - \frac{A_p^2B}{JV_p}v_c - \frac{A_r^2B}{JV_r}v_c - \beta_r^{\,(3)} +\\
&\underbrace{\frac{A_pB}{JV_p}Q_p(u) + \frac{A_rB}{JV_r}Q_r(u)}_{v_{\text{fm}} \text{(virtual control input) }}. 
\end{aligned}
\end{equation}

The virtual control input for FL is designed as:
\begin{equation}
v_{\text{fm}} = \frac{\dot{\hat{M}}_{d}}{J} + \frac{A_p^2 B}{JV_p}v_c + \frac{A_r^2 B}{JV_r}v_c + \beta_r^{\,(3)} - K_{\text{FL}} \underline{\mathbf{e}_\beta}
\end{equation}
to obtain the ideal exponentially stable dynamics $e_\beta^{(3)} = - K_{\text{FL}} \underline{\mathbf{e}_\beta}$,
with $\mathbf{e}_\beta = [e_\beta \; \dot{e}_\beta \; \ddot{e}_\beta ]^T$, $M_d = M_h + M_{Fr,b}$ and $\hat{M}_d$ being its estimate via disturbance observer. The same approach of \cite{dallabonaModelfreeSlidingMode2024} involving a state-augmented Luenberger observer, has been adopted for its design, resulting in bounded reconstruction errors $\delta_M, \dot{\delta}_M$ for both the disturbance and its derivative, respectively.

For the STSMC, a second-order sliding manifold is defined as:
\begin{equation}\label{eq:s}
s = \ddot{e}_\beta + \lambda_2\dot{e}_\beta + \lambda_1 e_\beta,
\end{equation}
where $\lambda_1,\lambda_2>0$ are tuning parameters. Its dynamics
\begin{equation}\label{eq:s_dot}
\dot{s} = e_\beta^{(3)} + \lambda_2\ddot{e}_\beta + \lambda_1 \dot{e}_\beta,
\end{equation}
depends on $u$ and thus $v$. The resulting control law is:
\begin{align}
v_{\text{st}} &= \frac{\dot{\hat{M}}_{d}}{J} + \frac{A_p^2 B}{JV_p}v_c + \frac{A_r^2 B}{JV_r}v_c + \beta_r^{\,(3)} - k_1 \lfloor s \rceil^{1 / 2} \nonumber\\
            &- k_2 \int_0^t \lfloor s(\tau) \rceil^0 \, d\tau \label{eq:v_st}
\end{align}
with $\lfloor s \rceil^{\upsilon} = \vert s\vert^\upsilon sgn(s)$, and $k_1,k_2>0$ selected according to the tuning rules in \cite{chalangaFiniteTimeStabilization2017}.

\subsection{Nonlinear Controller: Position–Force Cascade}

In the PFC design, the inner loop regulates the hydraulic force $F_h$, and the outer loop commands the required $F_{h,\text{ref}}$ to track the desired pitch angle. The force tracking error is defined as
\begin{equation}
e_F = F_h - F_{h,\text{ref}},
\end{equation}
where $F_{h,\text{ref}}$ is an appropriate force reference which will be defined later.

It is assumed that an exact estimate of its derivative is available, obtained using the 2nd-order sliding mode differentiator (SMD) defined in \cite{levantRobustExactDifferentiation1998} and implemented via the Toolbox proposed in \cite{andritschRobustExactDifferentiator2021}. The derivative reconstruction error enters the inner loop as a perturbation that vanishes in finite time. The force tracking error system has relative degree one:

\begin{equation} \label{eq:error_dynamics}
\dot{e}_F =  - \frac{A_p^2B}{V_p}v_c - \frac{A_r^2B}{V_r}v_c - \dot{F}_{h,\text{ref}} + \underbrace{\frac{A_pB}{V_p}Q_p(u) + \frac{A_rB}{V_r}Q_r(u)}_{v}.
\end{equation}
The virtual control law, i.e. the desired signal that $v$ needs to track, for both FL and STSMC is:

\begin{equation} \label{eq:psi_k}
v^* = v^*(t,e_F) = \frac{A_p^2B}{V_p}v_c + \frac{A_r^2B}{V_r}v_c + \dot{F}_{h,\text{ref}} - \psi_k(e_F)
\end{equation}
where the correction term $\psi_k$ is either of the following:
\begin{equation}
\psi_k(e_F) = 
\begin{cases}
K_{\text{FL}}e_F & \text{(FL)} \\
k_1 \lfloor e_F \rceil^{1/2} + k_2 \displaystyle \int_0^t \lfloor e_F(\tau) \rceil^0 \, d\tau & \text{(STSMC)}
\end{cases}
\end{equation}

For the outer loop, the pitch angle error dynamics from $F_{h,\text{ref}}$ to $\beta$ yields a second-order system:

\begin{equation}
\ddot{e}_\beta = \frac{(F_{h} - F_{Fr,c}) \,r_{\beta} - M_{Fr,b} - M_{\text{ext}}}{J} - \ddot{\beta}_r.
\end{equation}

Using feedback linearization, the control law, i.e. the desired hydraulic force becomes:
\begin{equation}
F_{h,\mathrm{ref}}(t,\underline{\mathbf{e}_\beta}) = F_{Fr,c} + \frac{J}{r_\beta}
\Biggl(
  \frac{\hat M_d}{J}
  + \ddot{\beta}_r
  - K_{\mathrm{FL,o}}
    \underbrace{
      \begin{bmatrix}
        e_\beta \\[4pt]
        \dot e_\beta
      \end{bmatrix}
    }_{\underline{\mathbf{e}_\beta}}
\Biggr),
\end{equation}
where $K_{\mathrm{FL,o}} = \begin{bmatrix}
    K_{\mathrm{FL,o},1} & K_{\mathrm{FL,o},2}
\end{bmatrix}$ is an appropriate gain matrix designed by any pole placement technique, and $\dot{e}_\beta$, $ \ddot{\beta}_r$ are also obtained using the SMD presented in \cite{levantRobustExactDifferentiation1998}, with bounded reconstruction errors $\delta_E$, $\delta_B$, respectively, that vanish in finite time. The P controller design reads $F_{h,\mathrm{ref}}(\underline{\mathbf{e}_\beta}) = -K_{P,o} \,e_{\beta}, \; K_{P,o} > 0$.

\subsection{Inversion Block}\label{sec:inversion}

The purpose of the inversion block is to invert the nonlinear mapping $u \mapsto \rho(t,u)$ defined by
\begin{equation}\label{eq:rho_hat}
v = \rho(t,u) = \frac{r_{\beta}A_pB}{J_{\text{tot}}V_p}Q_p(p_p,p_r,u) +  \frac{r_{\beta}A_rB}{J_{\text{tot}}V_r}Q_r(p_r,u),
\end{equation} 
and solve with respect to $u$ for a given $v$. Because $Q_p$ and $Q_r$ depend nonlinearly on $u$ via the pressure differentials $\delta_p,\delta_r$, solving for $u$ directly is generally intractable. A practical workaround, proposed in~\cite{ayalewCascadeTuningNonlinear2006}, consists in approximating $K(u)\approx K_uu$ and using the control input from the previous time step to estimate the current pressure drop, thereby breaking the algebraic loop. By defining
\begin{equation}
    g = \frac{r_{\beta}A_pB}{J_{\text{tot}}V_p}K_u\delta_p(p_p,u_{t-1}) +  \frac{r_{\beta}A_rB}{J_{\text{tot}}V_r}K_u\delta_r(p_r,u_{t-1}),
\end{equation}

the final control action is given by:
\begin{equation}
    u = \frac{v}{g}.
\end{equation}

An alternative solution was presented in \cite{dallabonaDynamicInputMapping2025}. This approach involves a dynamic mapping inversion (MI) algorithm, which introduces a differential mapping to solve for $u$ implicitly at each time step, thus preserving causality and avoiding the need for past-input approximations. The estimation is performed using the nonlinear estimator \cite{gripParameterEstimationCompensation2010b}, \cite{papageorgiouRobustBacklashEstimation2019}:
\begin{equation}\label{eq:estimation_law}
    \dot{u} = \gamma\,\mu(t,u) \, \Big( v - \rho(t,u) \Big),
\end{equation}
where $\gamma > 0$ is a tuning gain and $\mu(t,u)$ is a design-dependent weighting function. If $\mu(t,u)$ is chosen to satisfy the following properties for all pairs $(u^{*},u)$, the estimation law ensures that the origin of the estimation error $\tilde{u} \triangleq u^{*} - u$ is globally exponentially stable (GES) \cite{slotineAppliedNonlinearControl1991a}.
\begin{condition}\label{con:con1} \normalfont
    There exists a positive definite real function $\sigma(t,\boldsymbol{x})$ such that
    \begin{equation}
    	\mu(t,u^{*})\left.\frac{\partial \rho}{\partial u}\right|_u + \left. \frac{\partial \rho}{\partial u}\right|_{u^{*}} \mu(t,u) \geq 2\sigma(t,\boldsymbol{x}) ~.
    \end{equation}
\end{condition}
\begin{condition}\label{con:con2} \normalfont
    There exists a positive real number $L_{\mu} > 0$ such that
    \begin{equation}
    	\left \vert \rho(t,u^{*}) - \rho(t,u) \right \vert \leq L_{\mu} \vert \tilde{u} \vert \sqrt{\sigma(t)} \; .
    \end{equation}
\end{condition}
\begin{condition}[Persistency of excitation]\label{con:con3} \normalfont
    There exist positive constants $T,\alpha_0 > 0$ such that
    \begin{equation}
    	\int_t^{t + T} \sigma(\tau) d\tau \geq \alpha_0 \; .
     \end{equation}
\end{condition}

It has been shown that the selection of $\mu(t,u)$ as a modified gradient of $\rho$ \cite{dallabonaDynamicInputMapping2025} ensures that conditions \ref{con:con1}-\ref{con:con3} are satisfied.

\section{Stability analysis}\label{sec4}

The stability properties of the proposed nonlinear controllers are examined by organizing them into groups based on structural characteristics. Meanwhile, the PI‐based baseline controllers serve as a well‐understood benchmark, with their stability properties rigorously studied in the literature \cite{astromPIDControllersTheory1995}.

\subsection{Full Model controllers} \label{sec:full_model_stabilities}

This category comprises four controllers: the FL and ST designs, along with their dynamic mapping inversion counterparts, MIFL and MIST. In the MIFL design the state vector comprises the pitch tracking error along with its first and second derivatives, i.e. $\underline{\mathbf{e}_\beta}$ (defined in Section 3.1), as well as the input estimation error $\tilde{u}$ (defined in Section 3.3). 
The Uniform Ultimate Boundedness (UUB) of the closed-loop trajectories $( \underline{\mathbf{e}_\beta}(t), \tilde{u}(t) )$ in the MIFL case was proven in \cite{dallabonaDynamicInputMapping2025}, whereas the properties of plain FL and ST are well-studied in the literature and will be omitted here for brevity \cite{slotineAppliedNonlinearControl1991a}. In the case of MIST, arguments from \cite{loriaFeedbackCascadeinterconnectedSystems2008} will be invoked to prove uniform global asymptotic stability (UGAS) of the origin.

For a given virtual control law $v_{ST}$, consider the ideal input $u^*$ such that $v_{ST} = \rho(u^*)$. From Eqs. \eqref{eq:ddde} and \eqref{eq:v_st}, the sliding variable dynamics in closed loop reads:
\begin{equation}
\begin{aligned}
    \dot{s}   =& -k_1 \lfloor s \rceil^{1 / 2} - k_2 \int_0^t \lfloor s(\tau) \rceil^0 d \tau\\
    &+ \Delta(t) + \rho(t,u^*-\tilde{u}) - \rho(t,u^*),
\end{aligned}
\end{equation}
where the Lipschitz perturbation $\Delta$ accounts for cancellation mismatches, disturbance reconstruction and differentiation errors, \hl{as well as the effects of the eventual faults occurring in the system}. The additional term $\rho(t,u^*-\tilde{u}) - \rho(t,u^*)$ results from adding and subtracting $\rho(t,u^*)$ when expressing the closed-loop dynamics around the ideal input $u^*$.
For $s\neq 0$ one can define \cite{morenoStrictLyapunovFunctions2012}:
\begin{equation}
    \begin{aligned}
        &\zeta_1\triangleq \lfloor s \rceil^{1/2}\\
        &\zeta_2\triangleq -k_2 \int_0^t \lfloor s(\tau) \rceil^0 d \tau - \Delta(t)\\
        &\boldsymbol{\zeta} \triangleq \begin{bmatrix}
            \zeta_1 & \zeta_2
        \end{bmatrix}^T,
    \end{aligned}
\end{equation}
and rewrite the closed-loop system as:
\begin{equation}
    \boldsymbol{\dot{\zeta}} = \underbrace{\frac{1}{|\zeta_1|} \begin{bmatrix}
            -\displaystyle \frac{1}{2} k_1 & 1 \\
            -k_2 & 0
        \end{bmatrix} \boldsymbol{\zeta} + \begin{bmatrix}
            0 \\
            -d
        \end{bmatrix}}_{\displaystyle \boldsymbol{f_{\zeta}}(t,\boldsymbol{\zeta})} + \underbrace{\begin{bmatrix}
            -\displaystyle \frac{\rho(t,u^*-\tilde{u}) - \rho(t,u^*)}{|\zeta_1|} \\
            0
        \end{bmatrix}}_{\displaystyle \boldsymbol{h}(t,\tilde{u})},
\end{equation}
with $k_1, k_2$ selected to fulfill the conditions in \cite{chalangaFiniteTimeStabilization2017} and $d \triangleq \dot{\Delta}$. Since $\boldsymbol{h}(t,0)=\boldsymbol{0}$, the unperturbed system has a finite stable equilibrium point at the origin. This has been shown in \cite{morenoStrictLyapunovFunctions2012} along with the existence of an absolutely continuous positive definite scalar function $V_{ST} = \boldsymbol{\zeta}^TP_{ST}\,\boldsymbol{\zeta}$
with $P_{ST}$ being a positive definite matrix,  such that $\dot{V}_{ST}\leq -c||\boldsymbol{\zeta}||\triangleq -W_{ST}(||\boldsymbol{\zeta}||)$, where $W_{ST}(\cdot)$ is a class-$\mathcal{K}_\infty$ function and $c>0$.

\begin{figure}[b]
    \centering
    \includegraphics[width=0.8\columnwidth]{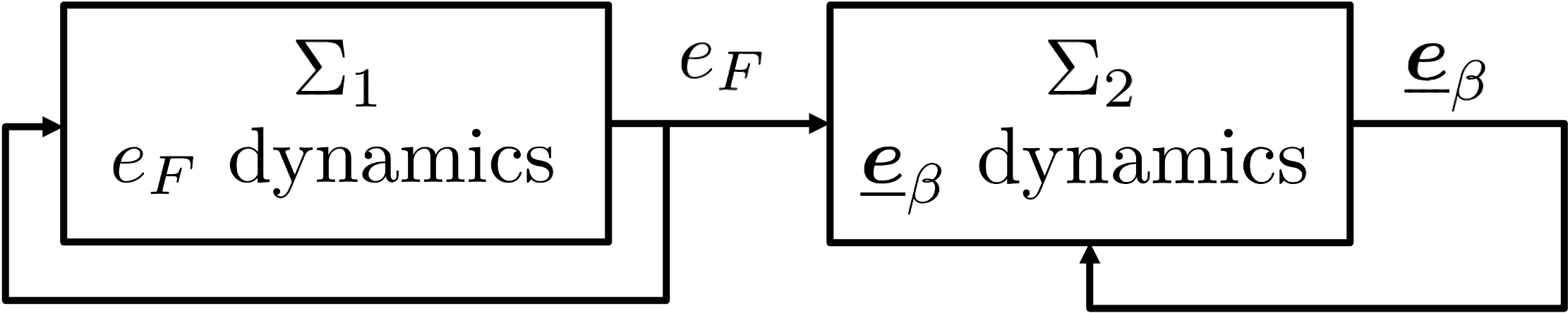}
    \caption{\hl{Cascaded interconnection of subsystems $\Sigma_1$ (hydraulic force tracking error) and $\Sigma_2$ (pitch tracking error) in the PFC architecture.}}
    \label{fig:caScadE}
\end{figure}

\begin{figure*}[ht]
    \centering
\includegraphics[width=0.85\textwidth]{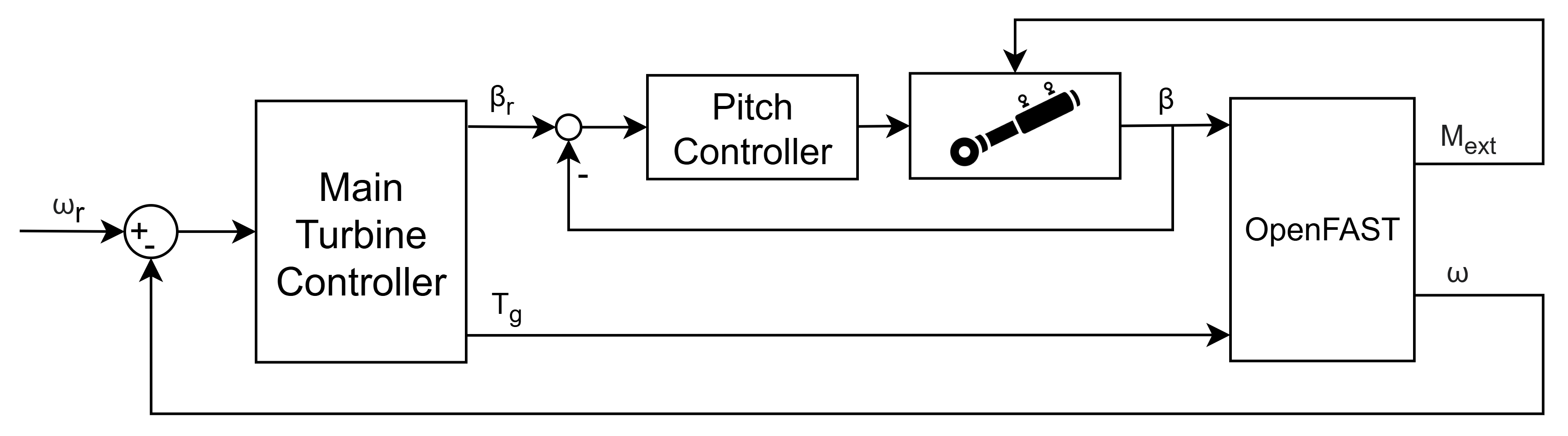}
    \caption{Block diagram of the OpenFAST-based simulation environment.}
    \label{fig:OF_sim}
\end{figure*}

The entire system dynamics can be seen as the feedback-interconnection of the sliding variable dynamics and the mapping inversion subsystem with their dynamics being written as:
\begin{equation}
    \begin{aligned}
        &\Sigma_{1,ST}: \dot{\boldsymbol{\zeta}} =  f_{\boldsymbol{\zeta}}(t,\boldsymbol{\zeta}) + \boldsymbol{h_{ST}}(t,\tilde{u})\\
        &\Sigma_{2,ST}: \dot{\tilde{u}} = \frac{d u^*}{dt} -\gamma \, \mu(t, \tilde{u})\, \Big( v_{st}(t,\boldsymbol{\zeta}) - \rho (u^* - \tilde{u}) \Big)
    \end{aligned}
\end{equation}
with
\begin{equation}
    \begin{aligned}
        f_{\boldsymbol{\zeta}}(t,\boldsymbol{\zeta}) &=  \frac{1}{|\zeta_1|} \begin{bmatrix} \displaystyle -\frac{k_1}{2} & 1 \\ -k_2 & 0 \end{bmatrix} \boldsymbol{\zeta} - \begin{bmatrix} 0 \\ d \end{bmatrix}\\
        \boldsymbol{h_{ST}}(t,\tilde{u}) &= -\begin{bmatrix} \displaystyle \frac{1}{|\zeta_1|} \\ 0 \end{bmatrix} \left( \rho(t,u^*-\tilde{u}) - \rho(t,u^*) \right),
    \end{aligned}
\end{equation}
where $v_{st}(t,\boldsymbol{\zeta})$ was defined in Eq.~\ref{eq:v_st}.
Since $v$ is a continuous signal (also in the case of the STSMC design) and the mapping $u \mapsto \rho(u)$ is a homeomorphism ($\partial \rho/\partial u \neq 0$) for $u \neq 0$
, the inverse $\rho^{-1}$ is well-defined and therefore implies differentiability of $u^{*}$ (see Appendix \ref{app:u_star_diff}).

Notice also that
\begin{enumerate}
    \item The unperturbed $\Sigma_{1,ST}$ $(\tilde{u}=0)$ is finite-time stable (FTS), which implies FTS of $s$ and, by extension, Uniform GES (UGES) of $\boldsymbol{e}_{\beta}$ \cite{slotineAppliedNonlinearControl1991a}.
    \item The subsystem $\Sigma_{2,ST}$ is UGES based on the selection of $\mu(t,\tilde{u})$ and conditions \ref{con:con1}-\ref{con:con3}.
    \item $\forall (t,\boldsymbol{\zeta}) \in [0,\infty) \times \mathbf{R}^2$ it holds that
    \begin{equation}
        \frac{\partial V_{ST}(t, \boldsymbol{\zeta})}{\partial \boldsymbol{\zeta}} \boldsymbol{h_{ST}}(t, \tilde{u}) = o(W_{ST}(\boldsymbol{\zeta})).
    \end{equation}
    The use of the smooth version of the Heaviside function defined in Eq.~\ref{eq:smooth_H} is reflected in the mapping $\rho(t, \tilde{u})$ being Lipschitz in $\tilde{u}$ with Lipschitz constant $L_{\rho}$ and
    \begin{equation}
    \begin{aligned}
        &\left\|\frac{\partial V_{ST}}{\partial \boldsymbol{\zeta}} \boldsymbol{h_{ST}}(t, \tilde{u}) \right\| 
        \leq \left\| 2 P_{ST} \boldsymbol{\zeta}\right\| \frac{L_\rho \|\tilde{u}\|_\infty}{\min_t(|\zeta_1(t)|)}  \\
        &\leq \underbrace{\left\| 2 P_{ST} \right\| \frac{L_\rho}{\min_t(|\zeta_1(t)|)} \|\tilde{u}\|_\infty }_{\eta'}  \left\|\boldsymbol{\zeta}\right\|.
    \end{aligned}
    \end{equation}
\end{enumerate}

Then by virtue of Proposition 1 in \cite{morenoStrictLyapunovFunctions2012}, the origin of the entire system is UGAS.

\subsection{PFC Controllers}

The closed-loop system resulting from each design based on the PFC architecture comprises two subsystems, namely the hydraulic force tracking error dynamics $\Sigma_1$ and the pitch tracking error dynamics $\Sigma_2$ as illustrated in Figure \ref{fig:caScadE}. The dynamics of these subsystems can be written in a general form as
\begin{align}
    (\Sigma_1): \dot{e}_F &= f_1(t,e_F) \label{eq:Sigma_1}\\ 
    (\Sigma_2): \underline{\dot{\mathbf{e}}_\beta} &= \boldsymbol{f_2}(t,\underline{\mathbf{e}_\beta}) + \boldsymbol{h_2}(e_F) \label{eq:Sigma_2}
\end{align}

Eq. \ref{eq:Sigma_2} has two variations, one corresponding to the P and FL controllers for the outer loop. Eq. \ref{eq:Sigma_1} has four: two (FL and ST) with static inversion and two (MIFL and MIST) with dynamic mapping inversion from Eq. \ref{eq:estimation_law}. In the latter case, Eq. \ref{eq:Sigma_1} takes the form
\begin{equation}\label{eq:Sigma_1_MI}
    \begin{matrix}
        (\Sigma_{11})\\
        (\Sigma_{12})
    \end{matrix}: \; \begin{bmatrix}
        \dot{e}_F\\
        \dot{\tilde{u}}
    \end{bmatrix} = \boldsymbol{f_1}(t,e_F,\tilde{u})
\end{equation}
since the dynamics of the input mapping inversion error must also be included.

Table \ref{tab:PFC_Subsystems} lists all possible variations of the dynamics of $\Sigma_1$ and $\Sigma_2$ along with the stability properties of the corresponding unperturbed subsystems. \hl{The terms $d_{f,\Sigma_1}$ and $d_{f,\Sigma_2}$ describe the time varying lumped effects of faults acting in the inner and outer loop, respectively.}
\begin{assum}\normalfont \label{assum:Boundedness_of_faults}
    \hl{The lumped fault perturbations $d_{f,\Sigma_1}(t)$ and $d_{f,\Sigma_2}(t)$ are uniformly bounded.}
\end{assum}
\begin{table*}[t]
    \centering
    \caption{Closed-loop dynamics and stability properties of subsystems in the PFC architecture.}
    \label{tab:PFC_Subsystems}
    \begin{tabular}{@{}llc@{}}
        \toprule
        \textbf{Inner loop} & $\boldsymbol{\Sigma_1}$ & Stability (unperturbed)\\
        \toprule
            FL & $\begin{aligned}f_1(e_F) &= -K_{FL}e_F -
            \delta_F + d_{f,\Sigma_1}
            \end{aligned}$ & UGUB\vspace{5pt}\\\hline
        ST & $\begin{aligned}f_1(e_F) &= -k_1 \lfloor e_F \rceil^{1/2} - k_2\int_0^t \lfloor e_F(\tau) \rceil^0 d\tau -
        \delta_F + d_{f,\Sigma_1}
            \end{aligned}$ & FTS\vspace{5pt}\\\hline
        MIFL & $\begin{aligned}\boldsymbol{f_1}(t,e_F,\tilde{u}) &= \begin{bmatrix}
            -K_{FL}e_F + v(t,u^* - \tilde{u}) - v^* -
            \delta_F + d_{f,\Sigma_1}\\
            \displaystyle \frac{d u^*}{dt} - \gamma \mu(t,\tilde{u})(v^*(t,e_F) - \rho(t,u^* - \tilde{u}))
        \end{bmatrix}
            \end{aligned}$ & UGUB\vspace{5pt}\\\hline
        MIST & $\begin{aligned}\boldsymbol{f_1}(t,e_F,\tilde{u}) &= \begin{bmatrix}
            -k_1 \lfloor e_F \rceil^{1/2} - \displaystyle k_2\int_0^t \lfloor e_F(\tau) \rceil^0 d\tau + v(t,u^* - \tilde{u}) - v^*
            - \delta_F + d_{f,\Sigma_1}\\\\
            \displaystyle \frac{d u^*}{dt} - \gamma \mu(t,\tilde{u})(v^*(t,e_F) - \rho(t,u^* - \tilde{u}))
        \end{bmatrix}
            \end{aligned}$ & UGAS\vspace{5pt}\\
        \toprule
        \textbf{Outer loop} & $\boldsymbol{\Sigma_2}$ & Stability (unperturbed)\\
        \toprule
        P & $\begin{aligned}\boldsymbol{f_2}(t,\underline{\mathbf{e}_\beta}) &= \begin{bmatrix}
                0 & 1\\
                \displaystyle -\frac{K_{P,o} r_{\beta}}{J} & \displaystyle - \frac{B_{v,b} }{J}
            \end{bmatrix}\underline{\mathbf{e}_\beta} - \begin{bmatrix}
                0\\
                \displaystyle \frac{r_{\beta}F_{Fr,c} + F_{c,b}\tanh(\gamma_L v_\beta) + B_{v,b}\,v_{\beta} + M_{ext}}{J} + \ddot{\beta}_r + d_{f,\Sigma_2}
            \end{bmatrix}\\
            \boldsymbol{h_2}(e_F) &= \begin{bmatrix}
                0\\
                \displaystyle \frac{1}{J}
            \end{bmatrix}e_F
        \end{aligned}$ & UGUB\vspace{5pt}\\\hline
        FL & $\begin{aligned}\boldsymbol{f_2}(t,\underline{\mathbf{e}_\beta}) &= \begin{bmatrix}
                0 & 1\\
                -K_{\mathrm{FL,o},1} & -K_{\mathrm{FL,o},2}
            \end{bmatrix}\underline{\mathbf{e}_\beta} - \begin{bmatrix}
                0\\
                \displaystyle \frac{\delta_M}{J} + \delta_E +
                \delta_B + d_{f,\Sigma_2}
            \end{bmatrix}\\
            \boldsymbol{h_2}(e_F) &= \begin{bmatrix}
                0\\
                \displaystyle \frac{1}{J}
            \end{bmatrix}e_F
        \end{aligned}$ & UGUB\vspace{5pt}\\
        \bottomrule
    \end{tabular}
\end{table*}
\hl{Considering the unperturbed inner loop ($\underline{\mathbf{e}_\beta} = \boldsymbol{0}$) the origin is at least UGUB for the FL and MIFL cases. Indeed, in the FL case, this is verified using $V_{FL} = \frac{1}{2}e_F^2$ as a Lyapunov function candidate and invoking the Input-to-State Stability (ISS) property of the system with respect to $d_{f,\Sigma_1}$ and Assumption }\ref{assum:Boundedness_of_faults}. Finite time stability (FTS)---and by extension UGAS---of the ST origin is well established in \cite{morenoStrictLyapunovFunctions2012} for appropriate selection of $k_1,k_2$. The case of MIFL has been studied \hl{for bounded perturbations} in \cite{dallabonaDynamicInputMapping2025} and finally UGAS for the MIST case can be shown using arguments identical to those in the proof in Section \ref{sec:full_model_stabilities}.

In the case of the unperturbed outer loop subsystems ($e_F = 0$), Uniform Global Ultimate Boundedness (UGUB) of the trajectories can be trivially shown since the system dynamics matrices - both for a P design and for a FL design - are Hurwitz and the external perturbations are uniformly globally bounded (UGB) \cite{slotineAppliedNonlinearControl1991a}. It should be noted that these perturbations also include the bounded and vanishing differentiation errors $\delta_E, \; \delta_B$ from the SMD used for estimating $\dot{e}_\beta, \; \ddot{\beta}_r$ and the bounded disturbance estimation error $\delta_M$.
\hl{Notice that the perturbations in $\Sigma_2$ are additive and the corresponding matrices are Hurwitz by design. Then it can be trivially shown that $\Sigma_2$ is ISS with respect to all bounded perturbations, including the solutions of $\Sigma_1$. It can therefore be concluded that the solutions of the entire cascaded interconnection are UGUB.}



\section{Simulation environment and experimental setup}\label{sec5}
The proposed control strategies have been evaluated in simulation using the OpenFAST framework~\cite{Jonkman2013}, which provides a high-fidelity nonlinear representation of wind turbine dynamics, integrated with the pitch system dynamics and the additional controllers required for closed-loop operation.

The pitch system model includes valve dynamics, rate limiters, and other relevant nonlinearities, and is parametrized to match the experimental pitch setup used for validation. The wind speed profile is modeled as a stochastic process with a mean value selected around the turbine’s rated wind speed, enabling the evaluation of the pitch controller performance both in the below-rated (MPPT) region--—where no pitching action is required—--and in the above-rated region, characterized by constant-power operation. Based on the measured wind speed, the \emph{Main turbine controller} is responsible for delivering the correct pitch reference $\beta_{r}$ to the system under analysis in the article.

\begin{table*}[tp]
    \centering
    \caption{Simulation parameters adopted for the pitch system, the proposed controllers, observer, and differentiators. 
    Exponential notation is written as $a^{\langle b\rangle}$ to indicate $a\times10^{b}$.}
    \label{tab:param}
    \begin{tabular}{@{}lcc@{}}
        \toprule
        \textbf{Parameter/System} & \textbf{Symbol (unit)} & \textbf{Value} \\
        \midrule

        \multicolumn{3}{@{}l}{\textbf{Pitch system parameters}} \\ 
        \midrule
        Moment of inertia (including blade) & $J$ (kg·m$^{2}$) & $2.0^{\langle4\rangle}$ \\
        Piston chamber area & $A_p$ (m$^{2}$) & $1.54^{\langle-2\rangle}$ \\
        Rod chamber area & $A_r$ (m$^{2}$) & $9.0^{\langle-3\rangle}$ \\
        Piston chamber initial volume & $V_{0,p}$ (m$^{3}$) & $1.37^{\langle-3\rangle}$ \\
        Rod chamber initial volume & $V_{0,r}$ (m$^{3}$) & $5.95^{\langle-4\rangle}$ \\
        Cylinder stroke & $x_{c,m}$ (m) & $1.314$ \\
        Bulk modulus & $B$ (Pa) & $1.4^{\langle9\rangle}$ \\
        Supply pressure & $p_\mathrm{s}$ (Pa) & $1.85^{\langle7\rangle}$ \\
        Return pressure & $p_\mathrm{t}$ (Pa) & $1.0^{\langle5\rangle}$ \\
        Bearing static friction torque ($v_\beta >0$) & $F_{c,b}$ (N·m) & $1.06^{\langle3\rangle}$ \\
        Bearing static friction torque ($v_\beta <0$) & $F_{c,b}$ (N·m) & $2.59^{\langle3\rangle}$ \\
        Static friction force (cylinder) & $F_{c,c}$ (N) & $2.6^{\langle3\rangle}$ \\
        Viscous friction coefficient (bearing) & $B_{v,b}$ (N·m·s·rad$^{-1}$) & $2.29^{\langle3\rangle}$ \\
        Viscous friction coefficient (cylinder) & $B_{v,c}$ (N·s·m$^{-1}$) & $6.5^{\langle3\rangle}$ \\
        Friction smoothing coefficient & $\gamma_t$ (-) & $5.0^{\langle3\rangle}$ \\
        Heaviside function smoothing coefficient & $\gamma_L$ (-) & $1.0^{\langle5\rangle}$ \\
        Natural frequency of valve dynamics & $\omega_v$ (rad·s$^{-1}$) & $2.4^{\langle2\rangle}$ \\
        Valve damping ratio & $\zeta_v$ (–) & $3.0$ \\
        Mechanical link length 1 & $l_1$ (m) & $1.25$ \\
        Mechanical link length 2 & $l_2$ (m) & $9.85^{\langle-1\rangle}$ \\
        Mechanical link length 3 & $l_3$ (m) & $1.87$ \\
        Offset angle & $\Phi_\mathrm{off}$ (rad) & $8.2^{\langle-1\rangle}$ \\
        \midrule
        \multicolumn{3}{@{}l}{\textbf{Controllers}} \\ 
        \midrule
        PI & $K_P,\, K_i$  & $4.0^{\langle1\rangle},\, 1.0^{\langle2\rangle}$ \\
        FL & $K_{FL,1},\, K_{FL,2},\, K_{FL,3}$  & $1.0^{\langle5\rangle},\, 6.5^{\langle3\rangle},\, 1.4^{\langle2\rangle}$ \\
        ST & $k_1,\, k_2,\, \lambda_1,\, \lambda_2$  & $3.4^{\langle1\rangle},\, 3.0^{\langle2\rangle},\, 2.0^{\langle3\rangle},\, 1.0^{\langle1\rangle}$ \\
        MIFL & $K_{FL,1},\, K_{FL,2},\, K_{FL,3},\, \gamma$  & $2.71^{\langle5\rangle},\, 1.47^{\langle4\rangle},\, 2.2^{\langle2\rangle},\, 8^{\langle-6\rangle}$ \\
        MIST & $k_1,\, k_2,\, \lambda_1,\, \lambda_2,\, \gamma$  & $1.4^{\langle1\rangle},\, 5.0^{\langle1\rangle},\, 4.0^{\langle3\rangle},\, 5,\, 5^{\langle-5\rangle}$ \\
        P–PI & $K_{P,o},\, K_P,\, K_i$  & $2.0^{\langle8\rangle},\, 4.0^{\langle-7\rangle},\, 4.0^{\langle-6\rangle}$ \\
        P–FL & $K_{P,o},\, K_{FL}$  & $2.5^{\langle6\rangle},\, 6.0^{\langle1\rangle}$ \\
        P–ST & $K_{P,o},\, k_1,\, k_2$  & $3.2^{\langle7\rangle},\, 6.9^{\langle3\rangle},\, 1.12^{\langle7\rangle}$ \\
        FL–FL & $K_{FL,o,1},\, K_{FL,o,2},\, K_{FL}$  & $7.84^{\langle2\rangle},\, 5.6^{\langle1\rangle},\, 1.4^{\langle2\rangle}$ \\
        FL–ST & $K_{FL,o,1},\, K_{FL,o,2},\, k_1,\, k_2$  & $7.84^{\langle2\rangle},\, 5.6^{\langle1\rangle},\, 7.98^{\langle3\rangle},\, 1.6^{\langle7\rangle}$ \\
        P–MIFL & $K_{P,o},\, K_{FL},\, \gamma$  & $1.0^{\langle8\rangle},\, 3.0^{\langle1\rangle},\, 5^{\langle-14\rangle}$ \\
        P–MIST & $K_{P,o},\, k_1,\, k_2,\, \gamma$  & $1.0^{\langle8\rangle},\, 3.45^{\langle3\rangle},\, 3.0^{\langle6\rangle},\, 5^{\langle-14\rangle}$ \\
        FL–MIFL & $K_{FL,o,1},\, K_{FL,o,2},\, K_{FL},\, \gamma$  & $9.0^{\langle2\rangle},\, 6.0^{\langle1\rangle},\, 1.0^{\langle2\rangle},\, 6^{\langle-14\rangle}$ \\
        FL–MIST & $K_{FL,o,1},\, K_{FL,o,2},\, k_1,\, k_2,\, \gamma$  & $3.6^{\langle4\rangle},\, 1.2^{\langle2\rangle},\, 3.45^{\langle3\rangle},\, 3.0^{\langle6\rangle},\, 5^{\langle-14\rangle}$ \\

        \midrule
        \multicolumn{3}{@{}l}{\textbf{Disturbance observer parameters}} \\ 
        \midrule
        Observer gains & $L_1,\, L_2$  & $4.0^{\langle1\rangle},\, -8.08^{\langle-6\rangle}$ \\

        \midrule
        \multicolumn{3}{@{}l}{\textbf{Exact differentiator parameters}} \\ 
        \midrule
        HOSMD $\dot{e}_\beta$ & $n,\, r$  & $2,\, 3.2^{\langle-1\rangle}$ \\
        HOSMD $\ddot{\beta}_r$ & $n,\, r$  & $3,\, 3.4$ \\
        HOSMD $\dot{F}_{h,\mathrm{ref}}$ & $n,\, r$  & $2,\, 3.5^{\langle3\rangle}$ \\

        \bottomrule
    \end{tabular}
\end{table*}

Although the stability analysis guarantees the existence of appropriate controller gains, their tuning---for each architecture---follows a heuristic approach based on the presented simulator. Multiple simulations with different tunings are performed, and the best-performing configuration is manually selected based on a trade-off between tracking accuracy and input smoothness. Manual tuning ensures that unstable or marginally stable configurations—which may appear optimal when evaluated solely via cost functions—are discarded in favor of robust and reliable solutions. \hl{More details are presented in the supplementary material}. The same tuning is applied both in simulation and during the experimental validation in the test setup. A schematic of the simulation environment is shown in Figure~\ref{fig:OF_sim}, and all the simulation and controller parameters are summarized in Tab.~\ref{tab:param}.

Since the calculation of $\beta_{r}$ is based on the turbine's operating conditions, and the performance of the controlled pitch actuation loop influences the overall turbine behavior, both the pitch reference and the aerodynamic loads depend on the controller under test. To enable fair experimental validation, both the pitch reference and wind loads are recorded during simulations for each controller configuration and replayed in the test rig.

\begin{figure}[t]
    \centering
    \includegraphics[width=0.96\columnwidth]{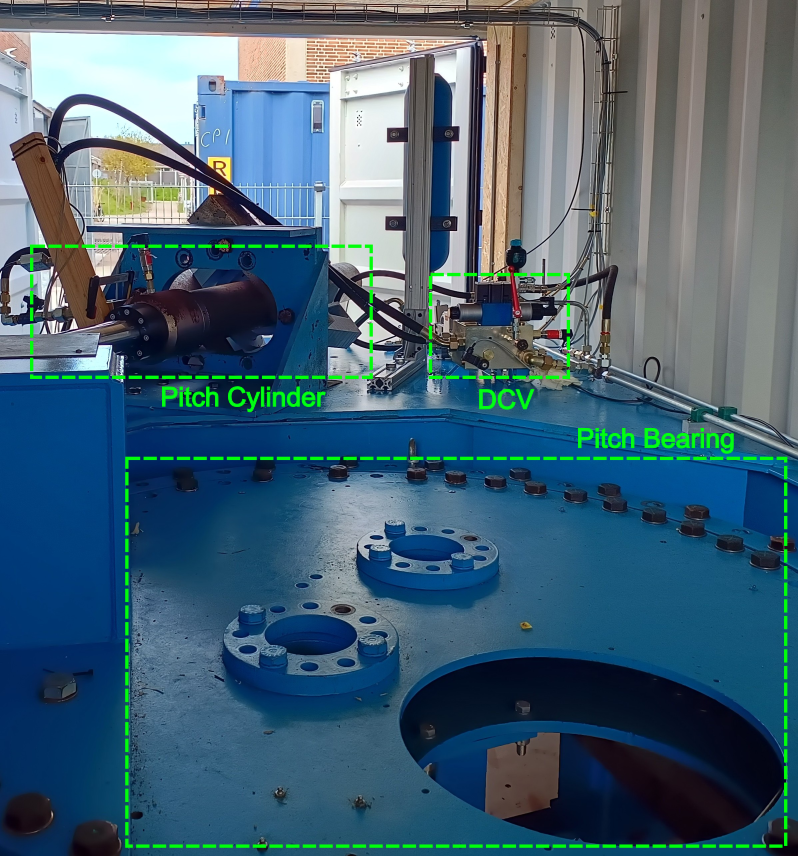}
    \caption{Experimental test rig for 3 MW turbine: commercial pitch actuator and blade mount supplemented with a dedicated hydraulic load emulator. Blade torque loads are calculated using time series from OpenFAST, that are converted to a time history of mechanical moments using the hydraulic load emulator. The load signal is compensated to emulate the full inertia of the turbine (Equation \ref{eq:blade_inertia_2}).}
    \label{fig:test_rig}
\end{figure}

The experimental setup is shown in Figure~\ref{fig:test_rig} and consists of a full-scale pitch system from a 3~MW commercial wind turbine. An additional hydraulic cylinder is used to replicate the aerodynamic loads recorded during the simulations. The setup does not include a physical blade for practical reasons, and is thus characterized by significantly reduced inertia.

To recreate realistic operational conditions, the missing blade inertia is compensated digitally. The real turbine equation of motion is given by:
\begin{equation} \label{eq:blade_inertia_1}
    (J_b + J_{\text{tr}})\,\dot{v}_\beta = M_h - M_{\text{fric}} - M_{\text{ext}}
\end{equation}

which can be equivalently expressed as:
\begin{equation} \label{eq:blade_inertia_2}
    J_{\text{tr}}\,\dot{v}_\beta \approx M_h - M_{\text{fric}} - \underbrace{\left(M_{\text{ext}} + J_b\,\hat{\dot{v}}_{\beta}\right)}_{M_{h,l}}.
\end{equation}

Here, $J_{\text{tr}}$ is the inertia of the test rig, $J_b$ is the (virtual) blade inertia, $M_h$ is the hydraulic actuation torque, and $M_{h,l}$ represents the load torque applied by the additional hydraulic cylinder.

Since direct computation of $\ddot{\beta}$ from measured signals is unreliable due to noise, the angular acceleration $\hat{\dot{v}}_{\beta}$ is obtained from simulation and replicated in the experimental setup via $M_{h,l}$.

All control algorithms are implemented and executed in real time using MATLAB/Simulink, deployed via a Speedgoat real-time controller with a sampling frequency of 1~kHz (1~ms time step).

\section{Results}\label{sec6}

Each control configuration is evaluated under nominal operating conditions as well as under the fault scenarios defined in Section~\ref{sec2}, to assess both nominal performance and passive fault rejection capabilities.

\hl{It should be noted that tests under faulty conditions assumed the occurrence of only one fault at a time. However, since the PFTC approach pertains to the robust rejection of the faults as perturbations, the proposed solutions can accommodate multiple faults happening simultaneously by simply lumping them into a single perturbation term in each loop.}

Results are presented in terms of selected performance metrics related to tracking fidelity and input signal quality. To isolate high-frequency oscillations in the control input $u$, a moving average $\bar{u}$ is computed over a fixed window. The filtered component is defined as $u_f = u - \bar{u}$, which has zero mean and only captures oscillatory content. The same filtering is applied to the valve position, yielding $x_{v,f}$.

The following performance metrics are used in the sequel:
\begin{multicols}{2}
\begin{itemize}
    \item \scalebox{0.85}{$\text{RMS}(u_f) = \sqrt{\frac{1}{N} \sum_{k=1}^{N} u_{fk}^2}$}
    \item \scalebox{0.85}{$\text{RMS}(x_{v,f}) = \sqrt{\frac{1}{N} \sum_{k=1}^{N} x_{v,fk}^2}$}
\end{itemize}
\begin{itemize}
    \item \scalebox{0.85}{$\text{RMS}(e_{\beta}) = \sqrt{\frac{1}{N} \sum_{k=1}^{N} e_{\beta,k}^2}$}
    \item \scalebox{0.85}{$\text{IAE}(e_{\beta}) = \int_0^T \vert e_{\beta}(t)\vert \, dt$}
\end{itemize}
\end{multicols}

where $T$ is the simulation duration. Finally\hl{, for visualization purposes, each metric value is normalized by its maximum value across the controllers} to facilitate meaningful comparisons across test cases.

\subsection{Nominal performance validation}

\begin{figure*}[tp]
    \centering
    \includegraphics[width=1.025\textwidth]{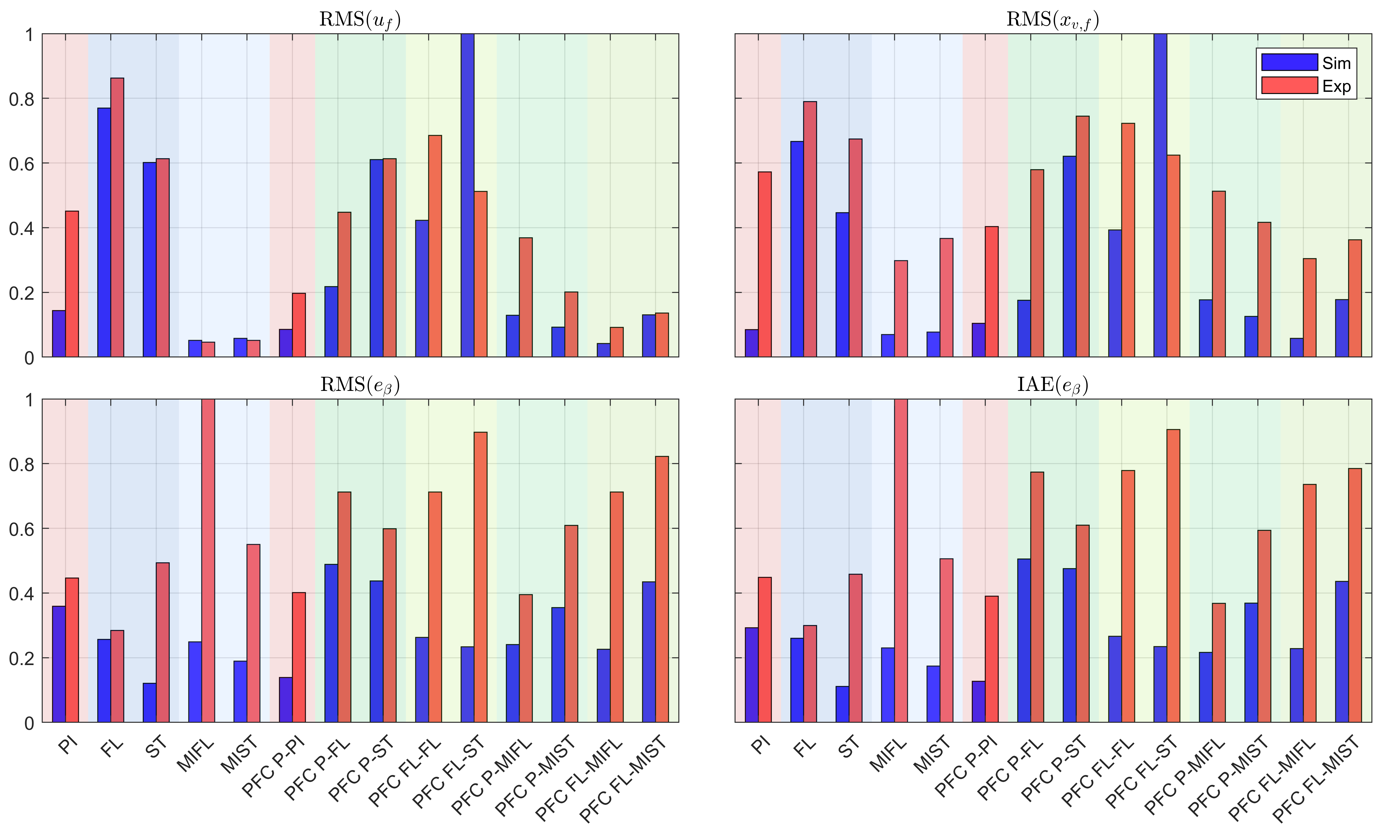}
    \caption{\hl{Performance results from simulation (blue) and experiment (orange) for the 14 different controllers.}}
    \label{fig:nom_res}
\end{figure*}

Figure~\ref{fig:nom_res} shows the nominal performance of each controller in simulation and experiment, normalized to 1. \hl{The results show broadly consistent qualitative trends, supporting the relevance of the simulation-based tuning. However, some quantitative discrepancies remain, which can be} attributable to:
\begin{itemize}
    \item Parametric and model uncertainty due to system wear and non-uniformity cause the experimental setup used in simulation to not replicate the test rig dynamics exactly;\\
    \item Virtual inertia reconstruction via estimated acceleration, which does not fully replicate real-life conditions \hl{and impacts the closed-performance of the setup};\\
    \item Non-ideal stiffness and position sensor measuring load cylinder displacement, leading to induced oscillations;\\
    \item Operation near the cylinder’s stroke limits, which introduces strong pressure gradients and increased sensitivity to model mismatch.
\end{itemize}

\begin{remark}
    Background shading in the plots denotes controller classes: red for PID-based solutions; blue (and light blue for MI variants) for full-model designs; green and yellow (with lighter shades for MI variants) for nonlinear PFC solutions with P (proportional) and FL outer controllers, respectively.
\end{remark}

Despite these factors, the following conclusions can be drawn:
\begin{enumerate}
    \item All controllers achieve bounded tracking errors using the same simulation-based tuning. The asymptotic convergence under nominal conditions is already reflected in the reported metrics; an explicit nominal time-domain tracking plot was therefore omitted to avoid redundancy\hl{---with representative tracking behaviour also illustrated under faulty conditions in Figure}~\ref{fig:fault_track}.\\
    \item The dynamic inversion algorithm substantially reduces chattering in both the control input and valve position, consistent across simulation and experiment. Consequently, previously highly-chattery nonlinear designs outperform the PI baseline in these metrics for both full-model and PFC-based cases.\\
    \item Although several controllers achieve lower normalized error metrics than the PI baseline in simulation, only a subset replicates these improvements experimentally, highlighting their significant potential, yet underscoring the challenges of implementing model‐based nonlinear designs on a real plant.
\end{enumerate}

\subsection{Fault tolerance performance validation}

The passive fault tolerant capabilities for most of the presented faults cannot be assessed experimentally, due to the impossibility of corrupting the test rig to replicate the most common faults affecting the real system. However, sensor faults can easily be reproduced digitally by corrupting the measurement before feeding it back in the loop. Since \hl{the $\beta$ sensor measurement is the only one utilized by the whole set of controllers,} the \hl{related} gain fault from Table~\ref{tab:fault_scenarios} was applied to a \hl{meaningful controller} selection. Again, because the pitch‐subsystem tracking performance influences the behavior of the entire wind-turbine control system, the pitch reference and wind load were recorded during the fault simulations for each selected controller. These profiles were then replayed on the test rig, yielding the most accurate experimental replication attainable.

\begin{figure}[t]
    \centering
    \includegraphics[width=0.98\columnwidth]{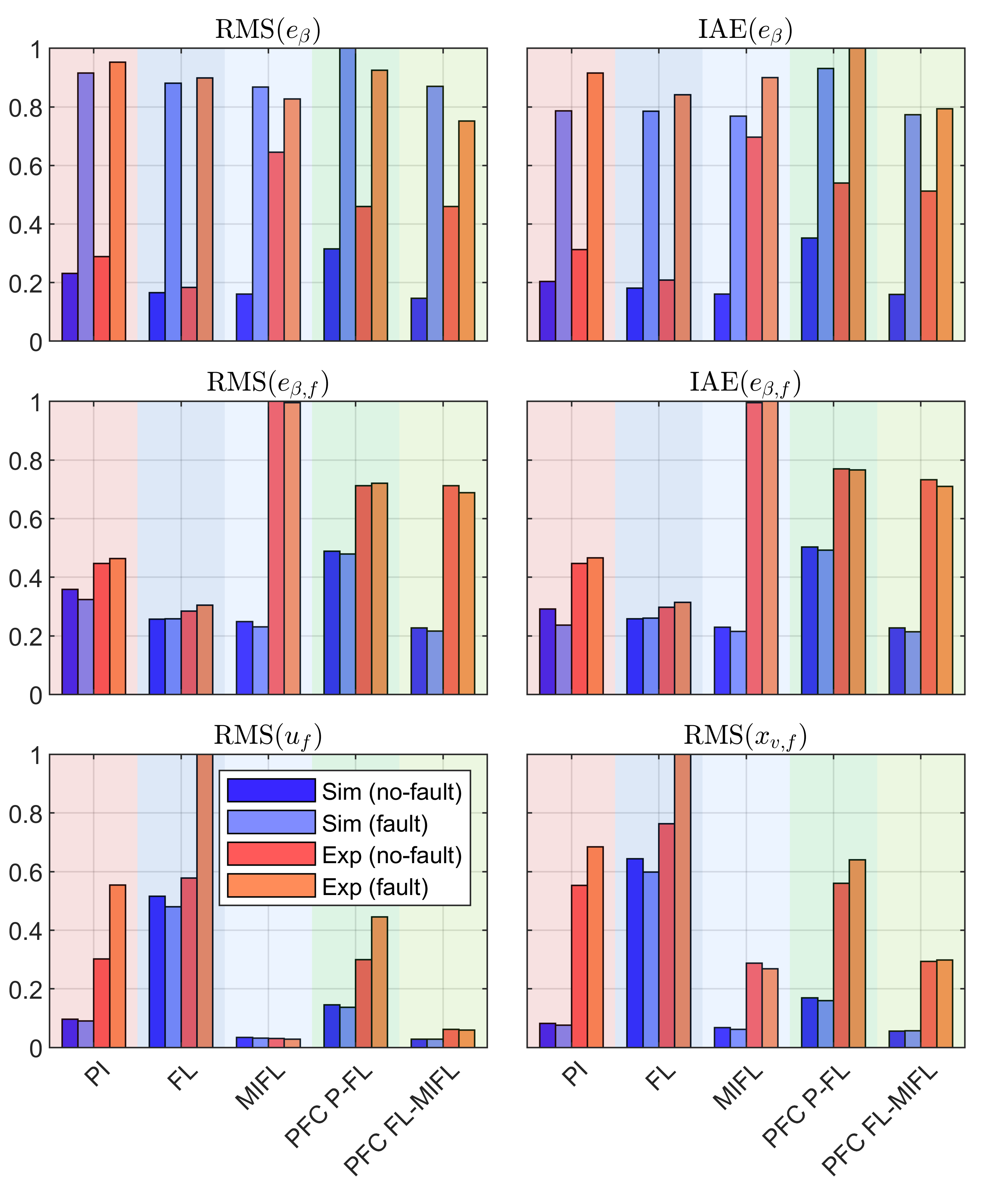}
    \caption{\hl{Sensor gain fault: Simulation and experiment, together with nominal case. Performance for the 5 main groups of controllers.}}
    \label{fig:fault_exp}
\end{figure}

The results are shown in Fig.~\ref{fig:fault_exp}, where tracking-error metrics are computed both from the actual pitch position, and from the sensor measurement corrupted by the fault (\(e_f\)). The former illustrates the pronounced impact of the fault, as the controller is unaware of the true pitch angle and thus behaves as though the reference signal were distorted. The latter indicates that, even under altered conditions (equivalent to a shifted reference), the apparent tracking performance differs little from the nominal case and is in some instances even improved. A more noticeable degradation appears in the chattering metrics, especially in the experiments.

\begin{figure}[t]
    \centering
    \includegraphics[width=0.93\columnwidth]{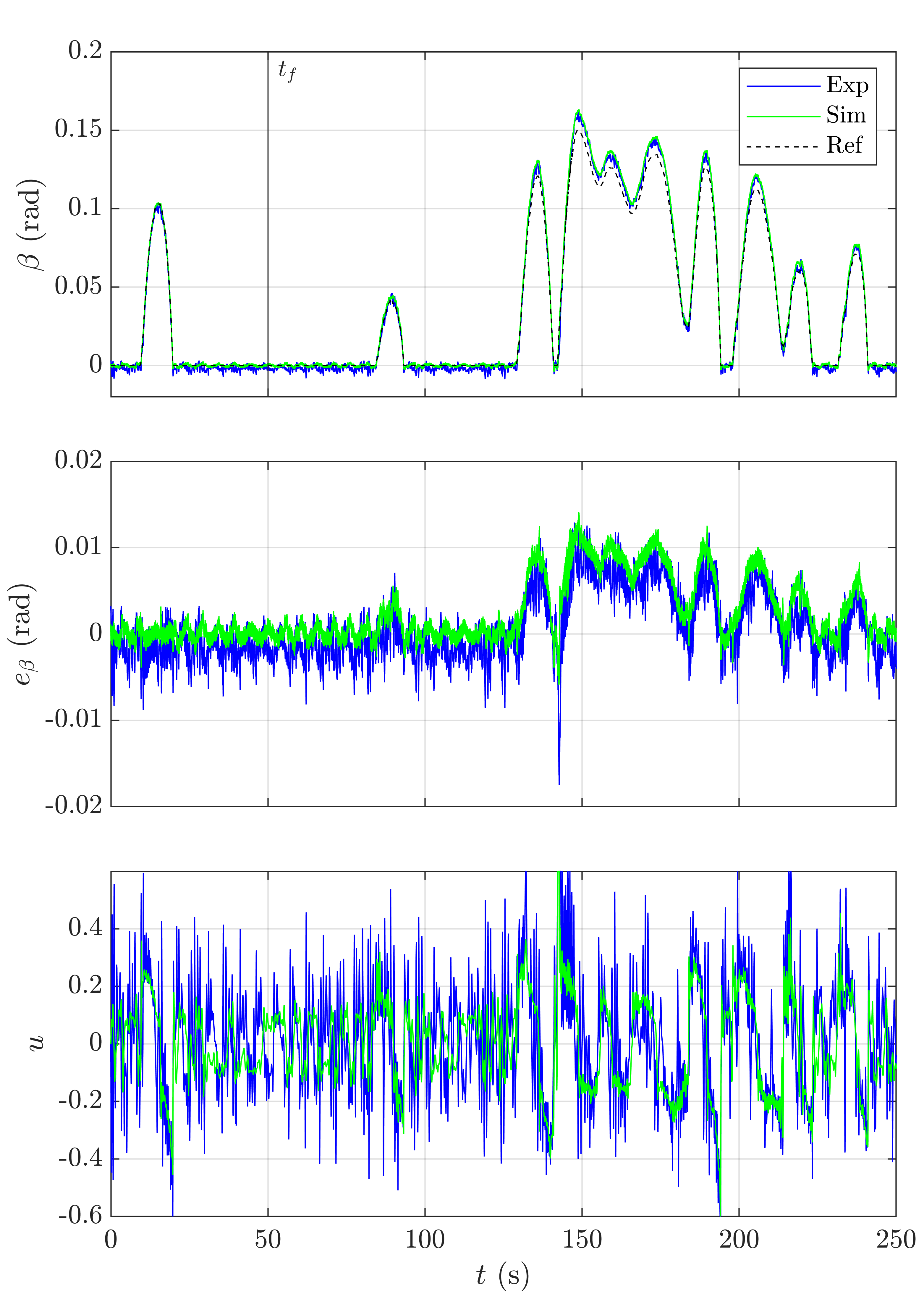}
    \caption{Sensor gain fault: Tracking performance example for FL-MIFL controller. Results from high fidelity simulation (green) and experiment (blue).}
    \label{fig:fault_track}
\end{figure}

\begin{figure}[h!]
    \centering
    \includegraphics[width=0.92\columnwidth]{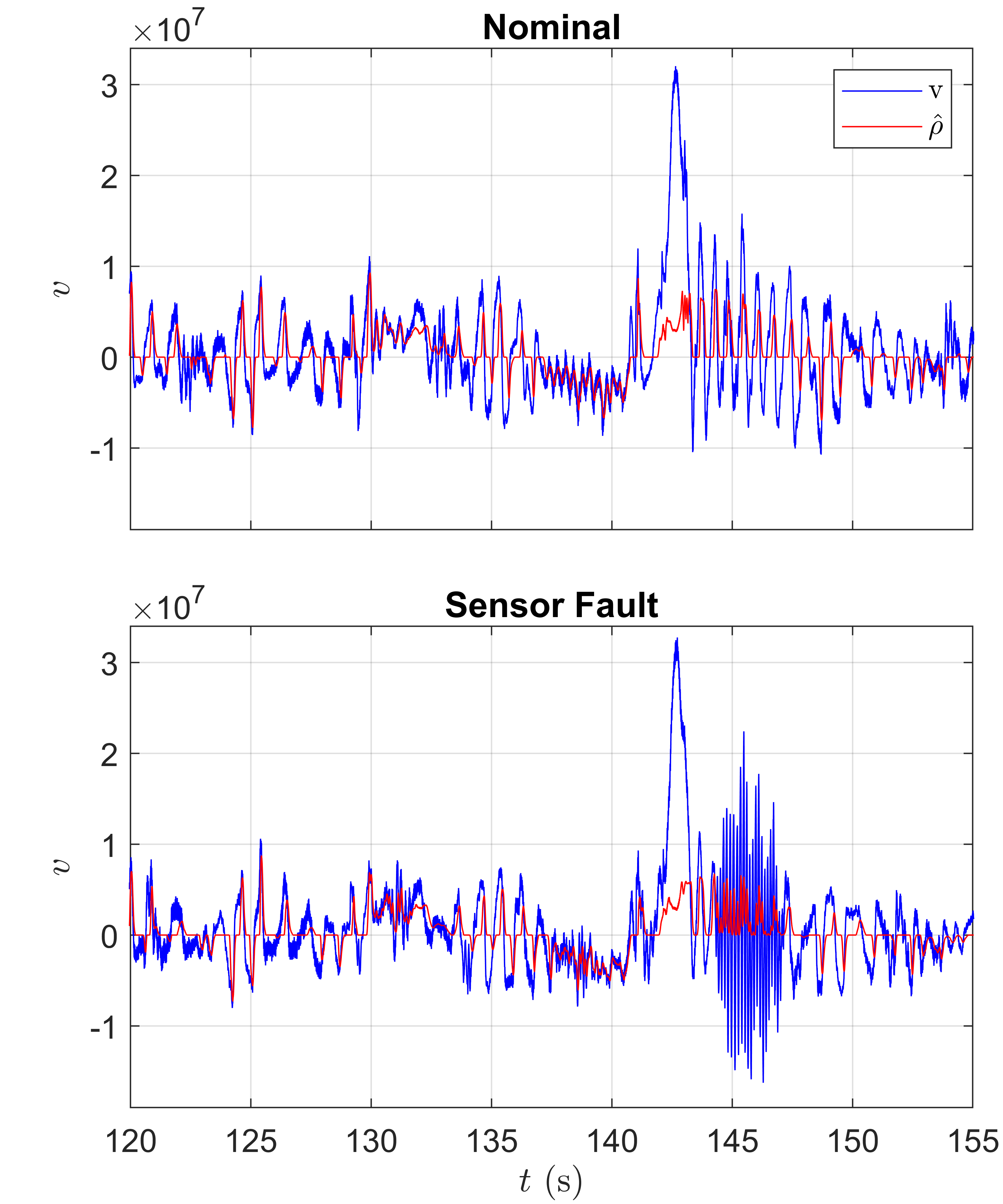}
    \caption{\hl{Sensor gain fault: Fault effect on mapping inversion algorithm for FL-MIFL controller in experiments.}}
    \label{fig:fault_mi}
\end{figure}

Figure~\ref{fig:fault_track} presents the FL–MIFL controller as a representative example, comparing simulation and experimental data. The plot depicts the tracking error calculated from the actual pitch angle, revealing a similar fault impact in both cases. Comparable behaviour is observed in the corresponding control input. According to Eq.~\ref{eq:rho_hat}, a faulty pitch reading also perturbs the internal variable \(x_c\) and thus the inversion algorithm. Figure~\ref{fig:fault_mi} presents the virtual-input reconstructions for FL–MIFL under nominal and faulty conditions, with the fault impact becoming evident at approximately 145~s.

\subsection{Complete fault tolerance analysis}

A systematic simulation study evaluates the \hl{individual} impact of every fault in Table~\ref{tab:fault_scenarios} on all the controllers \hl{that have been} validated experimentally. Whereas a single set of metrics is defined for each controller in the nominal case, each fault scenario produces its own metrics, so numerous plots would be needed for direct comparison. To condense the results, faults are grouped by category (Table~\ref{tab:fault_categories}) and a single composite metric is computed for each category. 
\hl{For this purpose, the metrics are normalized to their corresponding values obtained with the PI controller} (Table~\ref{tab:normalized_metrics})\hl{, setting the baseline to unity and facilitating comparisons across controllers and metrics. The normalized quantities are denoted, for example, by $\check{e}_{\mathrm{rms}}$.}

Error-related and input-related metrics are first averaged within each fault scenario \(i\)\hl{, with the latter also including $\mathrm{RMS}(u)$ to account for the overall control effort}:
\begin{equation}
\begin{aligned}
e_{\mathrm{eq},ij} &= \frac{1}{2}\bigl(\check{e}_{\mathrm{rms},ij} + \check{e}_{\mathrm{IAE},ij}\bigr),\\
u_{\mathrm{eq},ij} &= \frac{1}{3}\bigl(\check{u}_{\mathrm{rms},ij} + \check{u}_{f,\mathrm{rms},ij} + \check{x}_{v,f,\mathrm{rms},ij}\bigr).
\end{aligned}
\end{equation}
They are then averaged over the \(N_j\) faults in category \(j\):
\begin{equation}
e_{\mathrm{tot},j} = \frac{1}{N_j}\sum_{i=1}^{N_j} e_{\mathrm{eq},ij},\quad
u_{\mathrm{tot},j} = \frac{1}{N_j}\sum_{i=1}^{N_j} u_{\mathrm{eq},ij}.
\end{equation}
Finally, a combined score is defined as:
\begin{equation}
J_{\mathrm{tot},j} = \frac{1}{2}\bigl(e_{\mathrm{tot},j} + u_{\mathrm{tot},j}\bigr).
\end{equation}

\begin{table}[hb]
    \centering
    \begin{tabular}{@{}lc@{}}
        \toprule
        \textbf{Metric}     & \textbf{Value}                 \\
        \midrule
        $e_{rms,1}$         & $1.0539\times10^{-3}$         \\
        $e_{IAE,1}$         & $23.74$          \\
        $u_{rms,1}$         & $1.4479\times10^{-1}$         \\
        $u_{f,rms,1}$         & $7.7480\times10^{-3}$         \\
        $x_{vf,rms,1}$      & $3.1200\times10^{-3}$         \\
        \bottomrule
    \end{tabular}
    \vspace*{0.1cm}
    \caption{Normalized performance metrics for controller tracking and chattering (scientific notation)}
    \label{tab:normalized_metrics}
\end{table}

\begin{figure*}[h!]
    \centering    \includegraphics[width=0.95\textwidth]{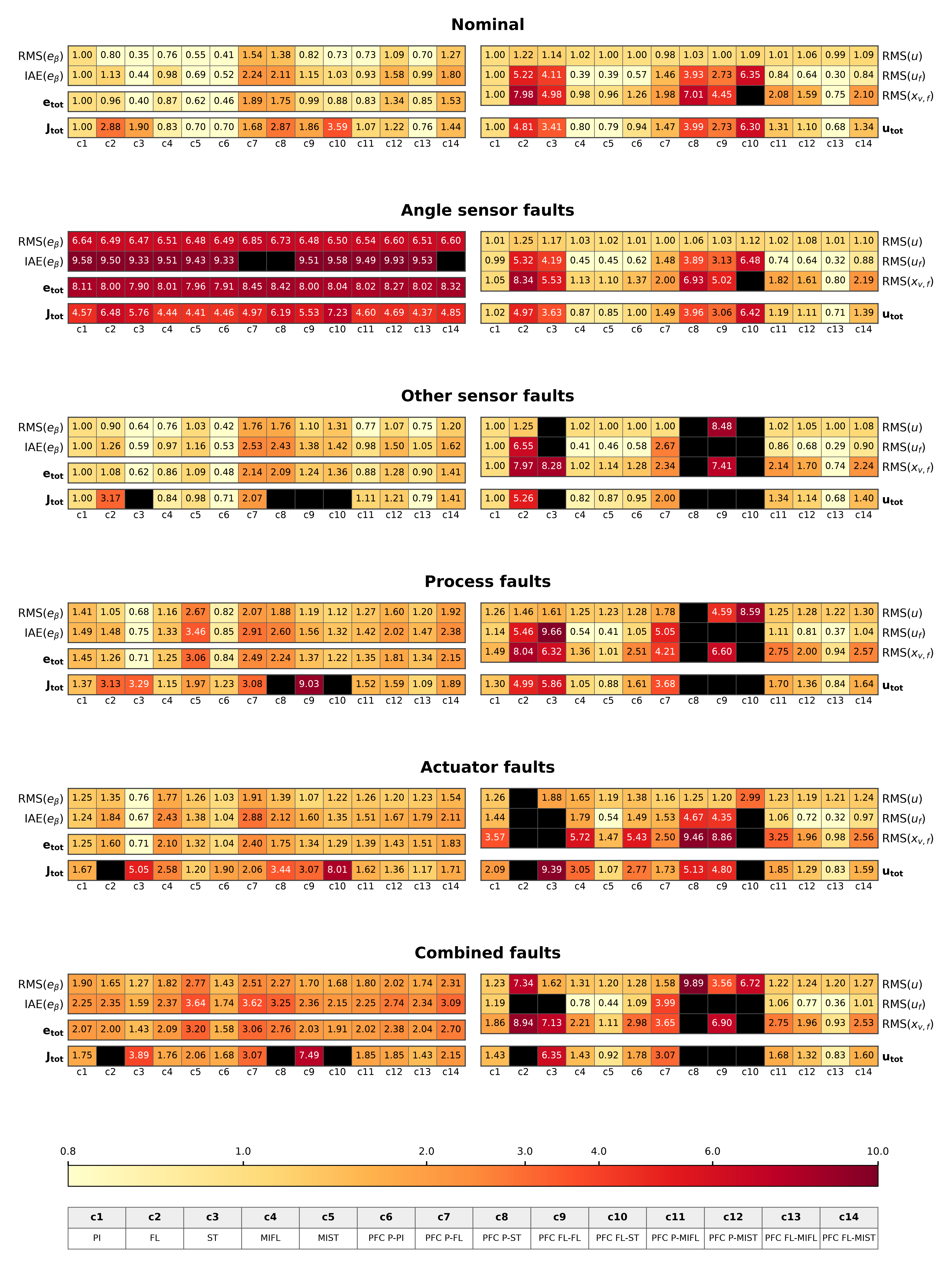}
    \caption{Fault tolerance scores for the proposed controllers under faulty conditions. The plot displays scores for tracking performance, input cleanliness and overall performance, for faults grouped in categories, with unity corresponding to the value of the metric for PI under nominal conditions.}
    \label{fig:fault_full}
\end{figure*}

An additional category aggregates all faults except “Other Sensor Faults”\hl{---as they do not impact all controllers---}yielding an overall fault-tolerance score. Figure~\ref{fig:fault_full} presents the normalized scores \hl{for both individual and grouped metrics}, with \hl{black cells} indicating values that exceed the plot range. The following conclusions can be drawn from this final analysis:

\begin{enumerate}
    \item Beta sensor faults have the strongest impact on the control action in terms of tracking error, as it is equivalent to tracking a different setpoint. The ratio in the metrics coefficient among the different controllers is similar to the other fault categories. An ad-hoc solution for actively compensating for such faults is essential for ensuring fault tolerance.
    \item Dynamic mapping inversion does not uniformly improve tracking in comparison to PI, but the significant chattering reduction renders the nonlinear designs practically implementable, compared to the standard counterparts characterized by unreasonable chattering levels, as shown by $u_{\mathrm{tot}}$.
    \item Full-model SMC achieves the best fault rejection in tracking error---according to $e_{\mathrm{tot}}$---due to its inherent robustness. The same doesn't apply when combined with PFC, where FL-based controllers exhibit better performance. On the other hand, the chattering level increases significantly, probably not justifying the slight rejection capabilities improvement.
    \item The FL–MIFL controller achieves the best overall passive fault-tolerance score $J_{\mathrm{tot}}$, with a normalized value close to unity, owing to its combined effectiveness in chattering reduction and fault rejection. In practical terms, its performance under fault conditions comes close to that of the PI benchmark in nominal operation. Other MI-enhanced designs approach this baseline, whereas non-MI designs generally yield higher (worse) scores.
\end{enumerate}

\section{Conclusion}\label{sec7}

A combined simulation–experimental study evaluated several nonlinear and robust strategies to obtain resilience against faults, wear and tear phenomena in wind turbine hydraulic pitch systems. Using identical simulation‐derived tuning, every controller tracked the reference accurately on the full‐scale test rig, confirming model fidelity. Dynamic mapping inversion (MI) proved pivotal by suppressing high-frequency oscillations in both the control input and valve position, and MI rendered the otherwise chattering‐prone nonlinear designs suitable for practical deployment.

Among the candidates, the FL–MIFL controller offered the best compromise. Its normalized fault-tolerance score remained close to unity, indicating that performance under faults is comparable to the PI baseline in nominal operation while preserving low chattering. Full-model sliding-mode designs exhibited the strongest tracking robustness but incurred unacceptable input oscillations unless combined with MI. Angle-sensor faults are shown to be the most disruptive ones\hl{, highlighting the limits of PFTC solutions and the need for targeted active ones.}

Overall, the study reveals an inherent trade-off between robustness and low actuator stress in passive fault-tolerant control, and identifies dynamic mapping inversion to be an effective means of reconciling these objectives. Future research will focus on integrating active fault-tolerant solutions that could complement and enhance the foundation established by the FL–MIFL baseline.

\section*{Acknowledgements}
This research was funded by The Danish Energy Technology Development and Demonstration Programme (EUDP) through the project: “Decreased Cost of Energy (CoE) from wind turbines by reducing pitch system faults”, grant number 64022-1058. The authors much appreciate this support. The authors would also like to thank Prof. H. C. Pedersen, as well as postdoctoral researchers M. Bhola and G. Wrat, and the students at Aalborg University, for their valuable assistance in conducting the experiments.

\appendices

\section{Proof of differentiability of $u^{*}$}
\label{app:u_star_diff}

Using the smooth approximation
\[
    H_{\gamma_L}(u^*) = \tfrac{1}{2}\bigl[1 + \tanh(\gamma_L\,u^*)\bigr],
\]
one obtains
\begin{align}
    \rho(u^*)
    &= k(u^*)\!\left[
        \sqrt{p_s - p_p}\,H_{\gamma_L}(u^*)
        + \sqrt{p_p - p_t}\,H_{\gamma_L}(-u^*)
    \right] \nonumber \\
    &= k(u^*)\bigl[c_0 + c_1 \tanh(\gamma_L\,u^*)\bigr]
    \,\triangleq\, k(u^*)\,g(u^*),
\end{align}
with
\begin{align}
    c_0 &= \tfrac{1}{2}\bigl(\sqrt{p_s - p_p} + \sqrt{p_p - p_t}\bigr),\\
    c_1 &= \tfrac{1}{2}\bigl(\sqrt{p_s - p_p} - \sqrt{p_p - p_t}\bigr).
\end{align}

For physically admissible operation of the valve, the chamber connected to tank
satisfies $p_p > p_t$, ensuring real-valued flow. This directly yields
\[
    c_0 - c_1 = \sqrt{p_p - p_t} > 0,
\]
and therefore $c_0 > c_1$. As a consequence, $g(u^*) > 0$ for all $u^*$. Moreover,
the valve flow coefficient $k(u^*)$ is monotonically increasing on $[-1,\,1]$,
where $u^*$ is defined, implying $k'(u^*) > 0$. Moreover,
\[
    g'(u^*) = c_1 \gamma_L \tanh(\gamma_L\,u^*)
\]
and both $k(u^*)$ and $g'(u^*)$ lie exclusively in the first and third quadrants, i.e.
$k(u^*)\,u^* > 0$ and $g'(u^*)\,u^* > 0$ for all $u^* \neq 0$. This implies
\begin{align}
    \operatorname{sgn}\,\!\bigl(k(u^*)\,g'(u^*)\bigr)
    &= \operatorname{sgn}\,\!\bigl(k(u^*)\bigr)
       \operatorname{sgn}\,\!\bigl(g'(u^*)\bigr) \nonumber \\
    &= \operatorname{sgn}\,\!\bigl(k(u^*)\,u^*\bigr) \;>\; 0,
\end{align}
for all $u^* \neq 0$. Therefore,
\[
    \rho'(u^*) \neq 0 \qquad \forall\, u^* \neq 0,
\]
and the mapping $u^* \mapsto \rho(u^*)$ is a global homeomorphism on the admissible
input domain, which is differentiable
everywhere except at $u^* = 0$, where no control action is applied. Consequently,
since the virtual control law $v$ is continuous—also in the STSMC case—the inverse
mapping exists, and the actual control signal $u^* = \rho^{-1}(v)$ is well defined and
differentiable for all $u^* \neq 0$.


\bibliographystyle{IEEEtran}
\bibliography{passive_comp_bib}


 




\vfill

\end{document}